\documentclass[%
 reprint,
amsmath,amssymb,aps,
]{revtex4-1}

\usepackage{soul} 
\usepackage{amsthm}
\usepackage{graphicx}
\usepackage{dcolumn}
\usepackage{bm}

\theoremstyle{definition}
\newtheorem{defn}{Definition}[section]

\begin{document}


\title{The impact of constrained rewiring on network structure and node dynamics}

\author{P. Rattana$^{1}$}
\author{L. Berthouze$^{2, 3}$}
\author{I. Z. Kiss$^{1}$}
\email{i.z.kiss@sussex.ac.uk}
\affiliation{$^1$ School of Mathematical and Physical Sciences, Department of
Mathematics, University of Sussex, Falmer, Brighton BN1 9QH, UK}

\affiliation{$^2$ Centre for Computational Neuroscience and Robotics, University of Sussex, Falmer, Brighton BN1 9QH, UK}
\affiliation{$^3$ Institute of Child Health, London, University College London, London WC1E 6BT, UK}

\date{\today}

\begin{abstract}
In this paper, we study an adaptive spatial network. We consider an $SIS$ (susceptible-infected-susceptible) epidemic on the network, with a link/contact rewiring process constrained by spatial proximity. In particular, we assume that susceptible nodes break links with infected nodes independently of distance, and reconnect at random to susceptible nodes available within a given radius. By systematically manipulating this radius we investigate the impact of rewiring on the structure of the network and characteristics of the epidemic. We adopt a step-by-step approach whereby we first study the impact of rewiring on the network structure in the absence of an epidemic, then with nodes assigned a disease status but without disease dynamics, and finally running network and epidemic dynamics simultaneously. In the case of no labelling and no epidemic dynamics, we provide both analytic and semi-analytic formulas for the value of clustering achieved in the network. Our results also show that the rewiring radius and the network's initial structure have a pronounced effect on the endemic equilibrium, with increasingly large rewiring radiuses yielding smaller disease prevalence.
\begin{description}
\item[Usage]
Secondary publications and information retrieval purposes.
\item[PACS numbers]
May be entered using the \verb+\pacs{#1}+ command.
\end{description}
\end{abstract}

\pacs{Valid PACS appear here}
\maketitle


\section{\label{sec:intro}Introduction}
The spread of infectious diseases on social networks and theoretical contact structures mimicking these has been the subject of much research~[\onlinecite{KeelingEames2005},\onlinecite{Danon2011},\onlinecite{Moreno2002},\onlinecite{Dorogstev2002}]. In general, most work in this area is aimed at understanding the impact of different network properties on how diseases invade, spread and how to best control them. Topological properties of nodes and edges can be exploited in order to minimise the impact of epidemics. For example, it is well known that isolating or immunising highly connected nodes or cutting edges or links with high betweenness centrality is far more efficient than selecting nodes and edges at random~[\onlinecite{Albert2000},\onlinecite{Holme2002}]. When global information is scarce, acquaintance immunisation~[\onlinecite{Cohen2003}] provides an effective way to significantly reduce the spread of an epidemic.
More recently, dynamic and time evolving network
models motivated by real data or simple empirical observations~[\onlinecite{Saramaki2005},\onlinecite{Gross2006},\onlinecite{Gross2008},\onlinecite{GrossBook},\onlinecite{Shaw2008},\onlinecite{Risau2009},\onlinecite{Lagorio2011}] have offered a different modelling perspective with important implications for how and when epidemics can spread or can be effectively controlled. It is widely accepted that during an epidemic the risk of becoming infected leads to social distancing with individuals either losing links or simply rewiring~[\onlinecite{Gross2006},\onlinecite{Sebastian2009},\onlinecite{Yilei2013},\onlinecite{Hatzopoulos2011}]. Such action can in fact be seen as an emerging control strategy.  In simple dynamic network models, contacts between susceptible and infectious individuals can be broken, and new ones be established. This is usually implemented by susceptible individuals breaking high risk contacts and rewiring to exclusively susceptible individuals or in a random way, or through random link addition and deletion~\cite{kissPRSA2012}. It has been shown that this adaptive mechanism has a strong impact on both epidemic dynamics and network structure.

Another major development is the consideration of spatial or geometric networks [\onlinecite{Barthelemy2011}], where nodes are embedded in space. This is especially the case for real networks where geographical or spatial location is key. For example, mobile phone, power grid, social contacts and neuronal networks are all embedded in space with location and proximity being a key component to how contacts are realised. This feature gives special properties to the network and allows to distinguish between nodes based on spatial proximity. For example, Dybiec et al.~\cite{Dybiec2004} proposed a modified $SIR$ (susceptible-infected-recovered) model using a local control strategy where nodes are distributed on a one dimensional ring, two-dimensional regular lattice and scale-free network. While infection could spread on the whole network, including shortcuts, control could only act over a `control network' composed of mainly local links but with neighbourhoods of varying size, e.g., including local neighbours one, two or more links away. They presented simulation results showing how the effectiveness of the local control strategy depends on neighbourhood size, and they explored this relationship for a variety of infection rates.

In order to make rewiring more realistic, it is possible to combine dynamic or adaptive networks with a spatial component, where nodes are given specific locations, such that the rewiring may take these locations into account when identifying  candidate nodes for rewiring. For example, Yu-Rong et al.~\cite{Song2013} considered a network with a spatial component, where the rewiring strategy was such that when an $SI$ link is cut, the $S$ individual will reconnect, with some probability $p$, to random individuals irrespective of distance, and to close-by or neighbouring individuals with  probability $1-p$. It was found that a higher value of the rewiring rate led to a lower final epidemic size whereas a smaller value of probability $p$ resulted in a slower epidemic spread.

In this study, we investigate an $SIS$ (susceptible-infected-susceptible) epidemic spreading on adaptive networks. Any susceptible node can avoid contact with infected
nodes by cutting its links to infectious nodes and by rewiring them to other susceptible nodes. However, we make the assumption that individuals may not be able to avoid connecting to individuals who are in the same community (e.g., social circles such as family, friends or workplace acquaintances). That is, whilst the network is rewired adaptively, the rewiring is restricted to susceptibles who are in the same `local' (to be defined later) area. The use of a square domain with periodic boundaries gives rise to a natural distance between nodes and this is used to determine the local area around nodes.

Since we anticipate that the size of local areas/neighbourhood will affect the rewiring, we carry out systematic numerical investigations of adaptive networks where rewiring is locally constrained. We adopt a step-by-step approach whereby we first study the impact of rewiring on the network structure in the absence of an epidemic, then with nodes assigned a disease status but without disease dynamics and, finally, running network and epidemic dynamics simultaneously. In the case of no labelling and no epidemic dynamics, we provide both analytic and semi-analytic formulas for the value of clustering achieved in the network in relation to the size of the local area.

The paper is structured as follows. In Section~\ref{sec:model}, we describe the construction of spatial networks to which constrained rewiring is applied, as well as the algorithm by which edges for rewiring are selected. We also present the impact of rewiring on degree distribution and clustering  when rewiring operates in the absence of an epidemic (Sections~\ref{sec:no-label}-\ref{sec:clustering}, respectively) and when the nodes are labelled (Section~\ref{sec:labelling}). Section~\ref{sec:full-model} describes the epidemic model with constrained rewiring, as well as numerical simulations of both homogeneous and heterogeneous networks. In Section~\ref{sec:discussion} we conclude the paper with a discussion of our results and possible further extensions of our work.

\section{\label{sec:model}Adaptive network model with locally-constrained rewiring}
In this section, the simplest adaptive network model with constrained rewiring is presented. Node placement and network construction are described by the following simple rules:

\begin{itemize}
\item $N$ nodes are placed uniformly at random on a square $S = [0,X] \times [0,Y] $, such that each node $i$ will have coordinates $0 \leq x_i \leq X$ and $0 \leq y_i \leq Y$, respectively, and $\forall i=1,2,\dots, N$.
\item Local area of radius $R$: If the Euclidian distance between nodes $i$ and $j$ is less than or equal to $R$, nodes $i$ and $j$ are said to be in the same local area, and can become connected during the rewiring process.
\end{itemize}

All results in this paper are derived by considering $S = [0,\sqrt{N}] \times [0,\sqrt{N}] $, and inter-nodal distances are calculated using periodic boundary conditions. With this choice, the density of nodes is exactly one node per unit area. Moreover, if the radius of the local area is $R$, then the circle, on average, will hold $n=\pi R^2$ nodes. Or if one wishes to control the expected number of nodes in a local area, then the radius is given by $R = \sqrt{n / \pi}$. Obviously, if $R\geq\sqrt{2N}/2$, the effect of spatial constraint
is non-existent as each node $i$ has $N-1$ potential neighbours to connect to. In what follows we will use either $n$, expected number of nodes in a local area, or $R$, the radius of that area as the control parameter of the rewiring process.

\subsection{\label{sec:no-label}Rewiring at random within local areas and impact of the local area radius}
We now investigate how changing the radius, which defines the local area for rewiring, affects the network structure. Here, in order to gain a better understanding of the rewiring algorithm, we study the network dynamics alone, in the absence of any dynamics of the nodes and without labelling nodes. Starting from the original idea of
cutting a link between a susceptible node $S$ and an infectious node $I$, and rewiring the susceptible to another $S$ node randomly chosen among the set of all susceptible nodes~\cite{Gross2006}, we consider two scenarios for implementing locally constrained rewiring. Specifically, we explore two different edge selection mechanisms:
\begin{enumerate}
\item {\em link-based selection:} a $SI$ link is chosen at random (with equal probability), after which, the susceptible node $S$ in the link is rewired to a randomly chosen available susceptible node $S$;
\item {\em node-based selection:} a susceptible node $S$ is chosen at random and, if connected to an infectious node $I$, is rewired to a randomly chosen available susceptible $S$.
\end{enumerate}

Unlike the node-based selection mechanism, the link-based selection mechanism favours highly-connected nodes and therefore these two selection mechanisms have the potential to lead to networks with different properties. Note that, in both cases, once a prospective link or node has been identified, rewiring happens according to the local constraint, that is, rewiring happens only if  at least one susceptible node $S$ is available in the local area. Otherwise, rewiring is not performed. The total number of edges is kept constant throughout the simulations, and rewiring is not allowed if it leads to self-connections or multiple/repeat
connections.

To begin to consider the impact of the network dynamics and show how it depends on the choice of selection algorithm and size of local area, we consider two different starting conditions: (a) homogeneous and (b) heterogeneous Erd\H{o}s-R\'enyi networks with average connectivity $\langle k \rangle$ = 10. Then, when $R$ =
$\sqrt{2N}/2$ or $n = N$, the network will be in the situation where $\langle k \rangle \ll n$, whereas when $R$ =
$\sqrt{6/\pi}$, we will have $\langle k \rangle \gg n$. 
In one simulation step, only two outcomes are possible: the rewiring is successful (one link has been cut and a new `local' link has been created) or the rewiring fails, as there are no suitable nodes in the local area. The latter tends to be more likely when the number of nodes in the local area is close to, or smaller than, average connectivity, as this means that after a few successful steps, new links would lead to multiple or repeat connections, which are not allowed. The simulations or rewiring steps are performed until network characteristics such as degree distribution and clustering have stabilised.

\begin{figure}[h!]
\includegraphics[scale=0.28]{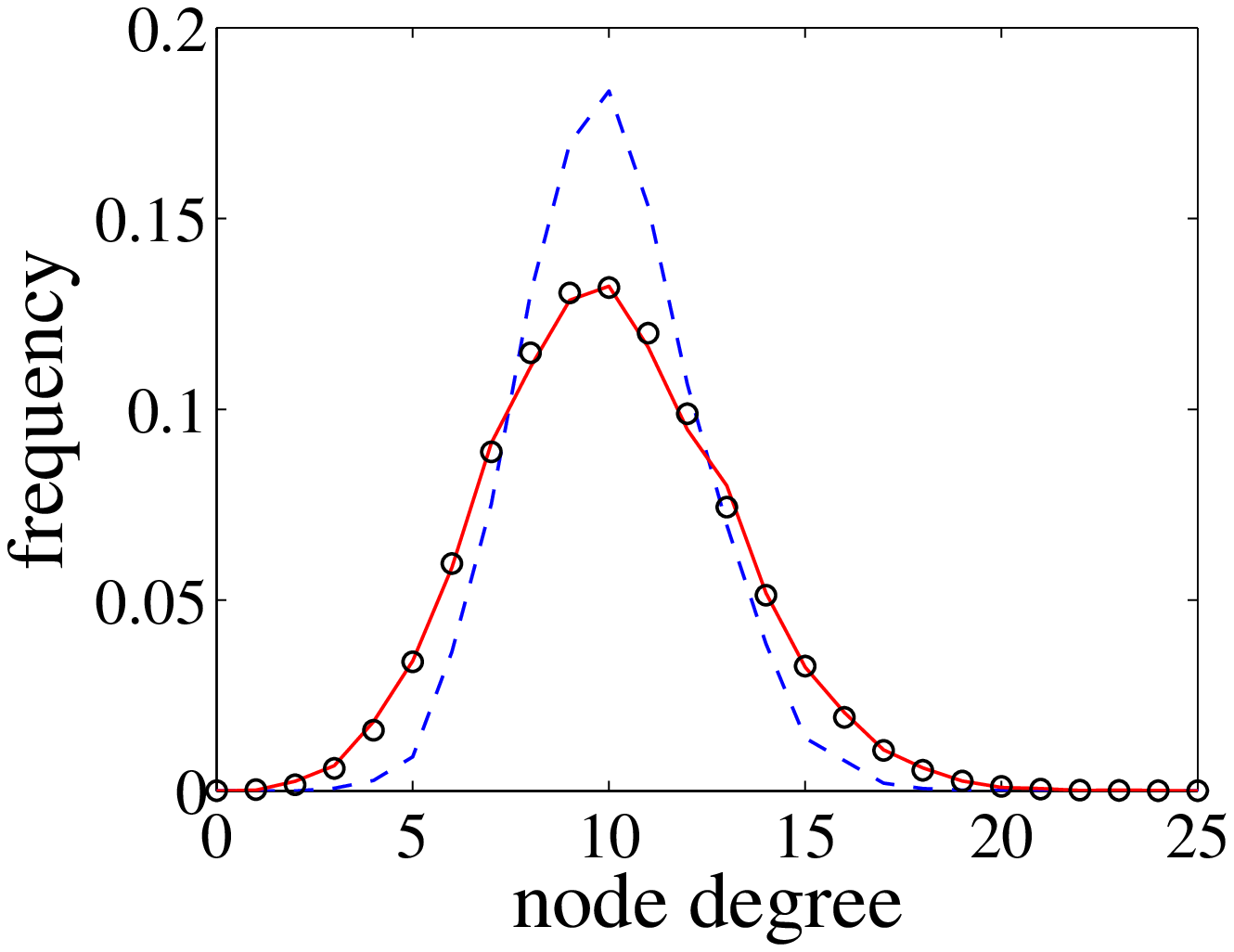}
\includegraphics[scale=0.28]{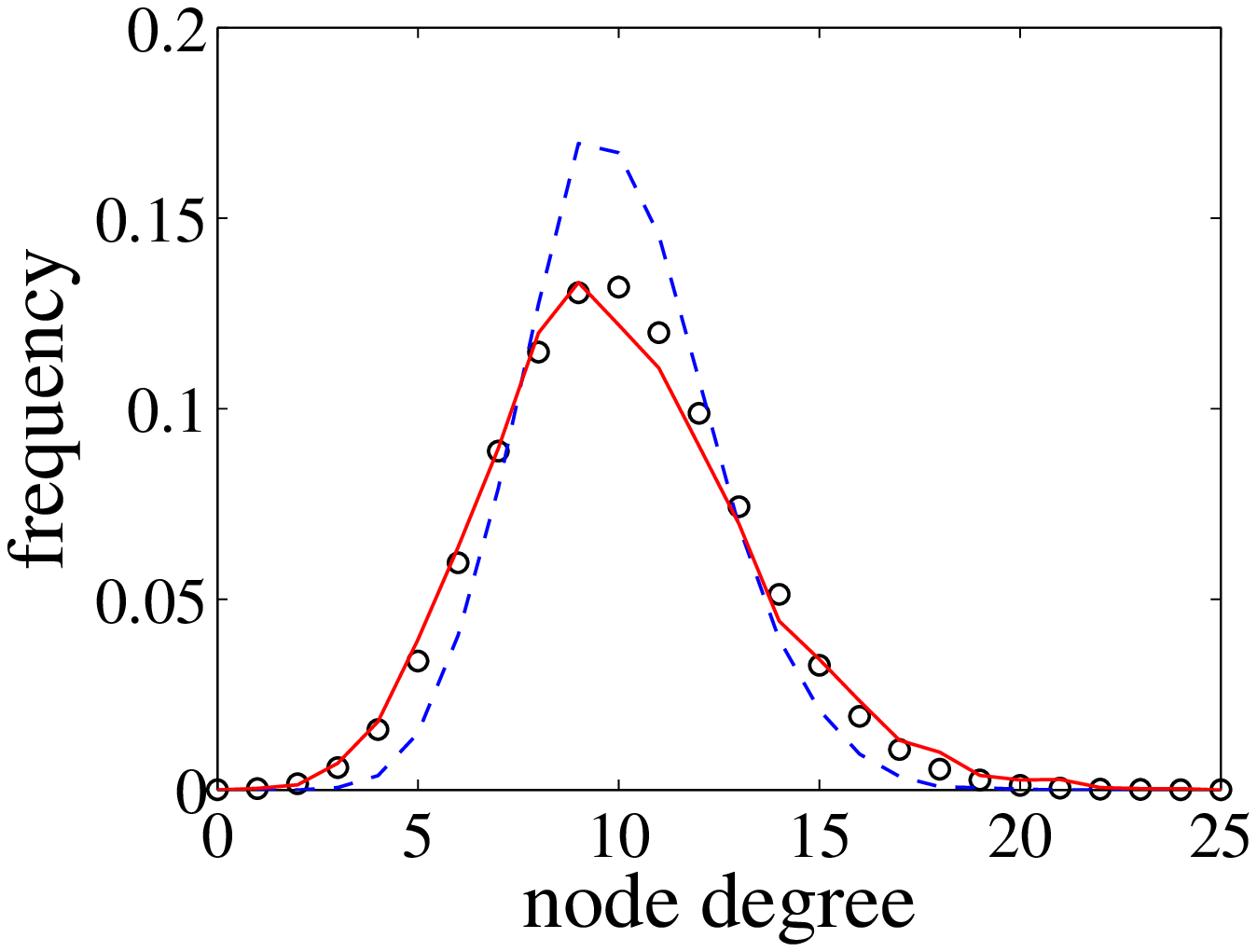}
\includegraphics[scale=0.28]{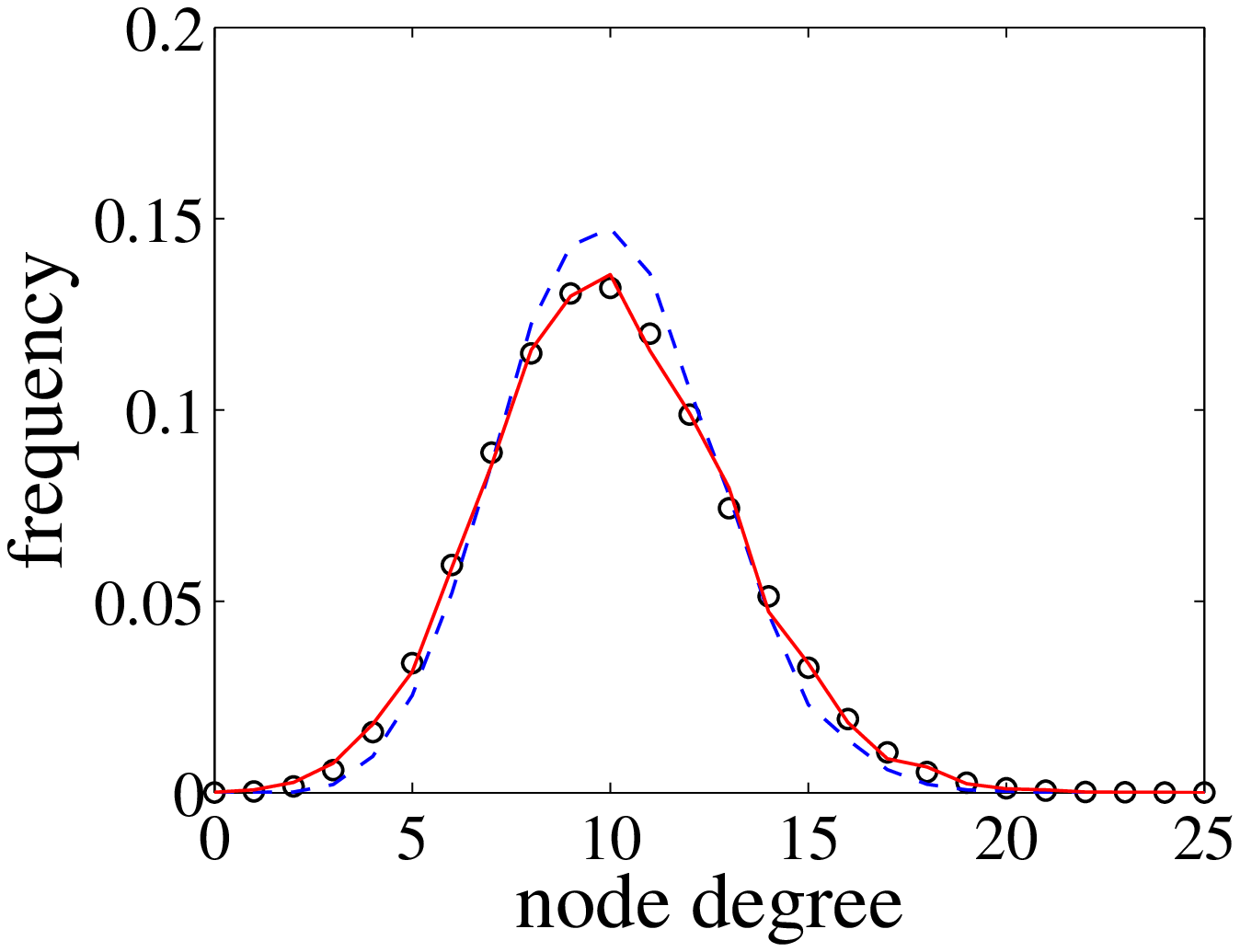}
\includegraphics[scale=0.28]{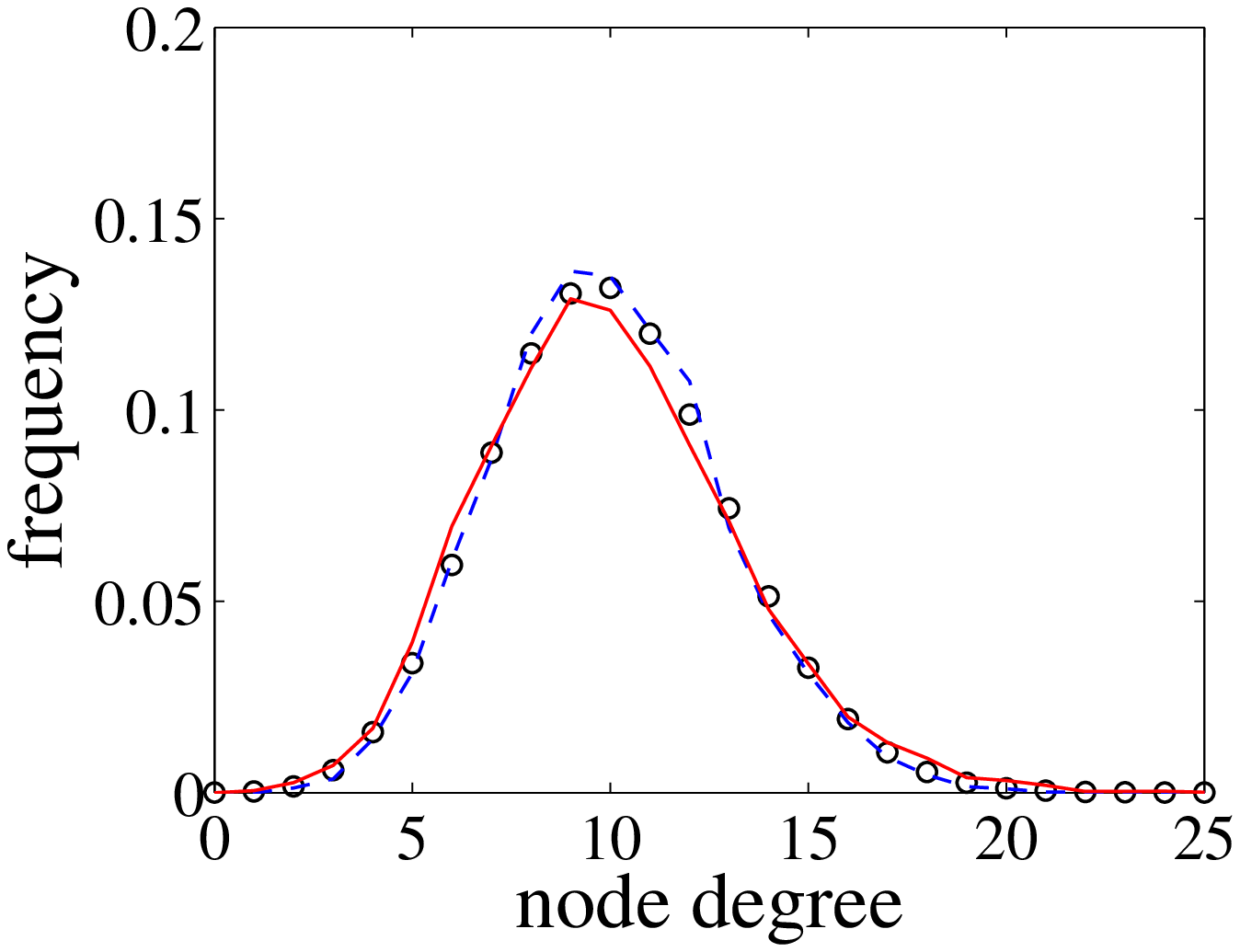}
\caption{\label{fig-10k-distribution} The average degree distribution at the end of simulations starting from homogeneous (top) and heterogeneous (bottom) networks compared with the binomial distribution $X \sim B(N-1,\langle k \rangle/(N-1))$
(black circles, corresponding to an Erd\H{o}s-R\'enyi random network with $N$ nodes and connectivity $\langle k \rangle$). The left and right panels correspond to link- and node-based selection, respectively.
The plots show the average of 100 simulations with $R$ = $\sqrt{2N}/2$ (red line) and $R$
= $\sqrt{6/\pi}$ (blue dash line), with $N$ = 100 and $\langle k \rangle$ = 10.}
\end{figure}

Fig.~\ref{fig-10k-distribution} shows the average or expected degree distribution at steady state for both
link-based and node-based selection methods. The good agreement between simulation and binomial distribution, when $R$ = $\sqrt{2N}/2$, confirms that the degree distribution has not changed for the random network, but has changed significantly for homogeneous network with both selection methods leading to a heterogeneous network.

Starting from homogeneous and heterogeneous networks leads to different outcomes, with the difference most pronounced at the peak of the degree distribution when $R$ =
$\sqrt{6/\pi}$. Namely, the peak of the degree distribution when using link-based selection is higher than that obtained when using node-based selection, and the peak when starting from heterogeneous networks is less than that starting from homogeneous network. These differences can be explained as follows.

For small local areas, where the average number of nodes is smaller than the average degree or connectivity, the rewiring will not be able to rewire all original links such that the final/stable distribution remains relatively close to the original or starting distribution. Hence, starting with a homogenous network with distribution $p(k)=\delta({k-\langle k \rangle})$, i.e. $p(\langle k \rangle)=1$, will lead to a network with a distribution that will maintain a high peak around $\langle k \rangle$. The heterogenous network has a much lower peak to start with, namely $p(\langle k \rangle)={N-1 \choose \langle k \rangle} p^{\langle k \rangle} (1-p)^{N-1-\langle k \rangle}$, where $p=\langle k \rangle/(N-1)$, and thus further limited rewiring will flatten the distribution further.

A similar explanation holds for the difference in peak when the starting network is the same but the selection method differs. This is a result of the selection algorithm, and we will consider the case when the starting network is homogenous. Some nodes with connectivity higher than $k$ will emerge quickly and these will be favourably picked for rewiring when the link-based algorithm is used. However, this will only lead to conserving the nodes' degree, and rewiring will only lead to an increase in the maximal
degree in the network if the target of the rewiring is itself one of the already highly connected nodes. This becomes very limiting and leads to little growth in degree, and thus to limited flattening of the distribution or decrease in its peak. This is exacerbated when the rewiring is limited by fewer available nodes than the
average connectivity.

The size of the local area has a significant effect on the number of nodes in the area. If we consider small values of $R$, such as $R$ = $\sqrt{6/\pi}$ and $\langle k \rangle > n$, then a typical node will connect to almost all nodes within the local area during the rewiring process. In other words, while the rewiring process is happening, the small number of nodes in the area will become well connected and will lead to the formation of triangles, and thus increasing levels of clustering. In the extreme case with only three nodes in the local area, a triangle will quickly form. When the average connectivity is similar to the number of nodes in a local area, the rewiring process will create a significant number of closed loops of length three, which will have a significant impact on the spread of a disease. To quantify this effect in a more rigorous way, we measure clustering in the network for local areas of different sizes as well as its evolution in time. Clustering can be simply calculated as the ratio of triangles to connected triples, open or closed. This can be computed by simple operations on the adjacency matrix of the network as follows:
\begin{equation*}
    C =  \displaystyle{\frac{\text{number of triangles}}{\text{number of triples}}} = \displaystyle{\frac{\text{trace}(G^3)}{\|G^2\|-\text{trace}(G^2)}},
\end{equation*}
where $G=(g_{ij})_{i,j=1,2,...N}\in{\{0,1\}}^{N^{2}}$ and
$g_{ij}=1$ if there is a connection between node $i$ and node $j$
and $g_{ij}=0$ otherwise.

\begin{figure}[h!]
\includegraphics[scale=0.28]{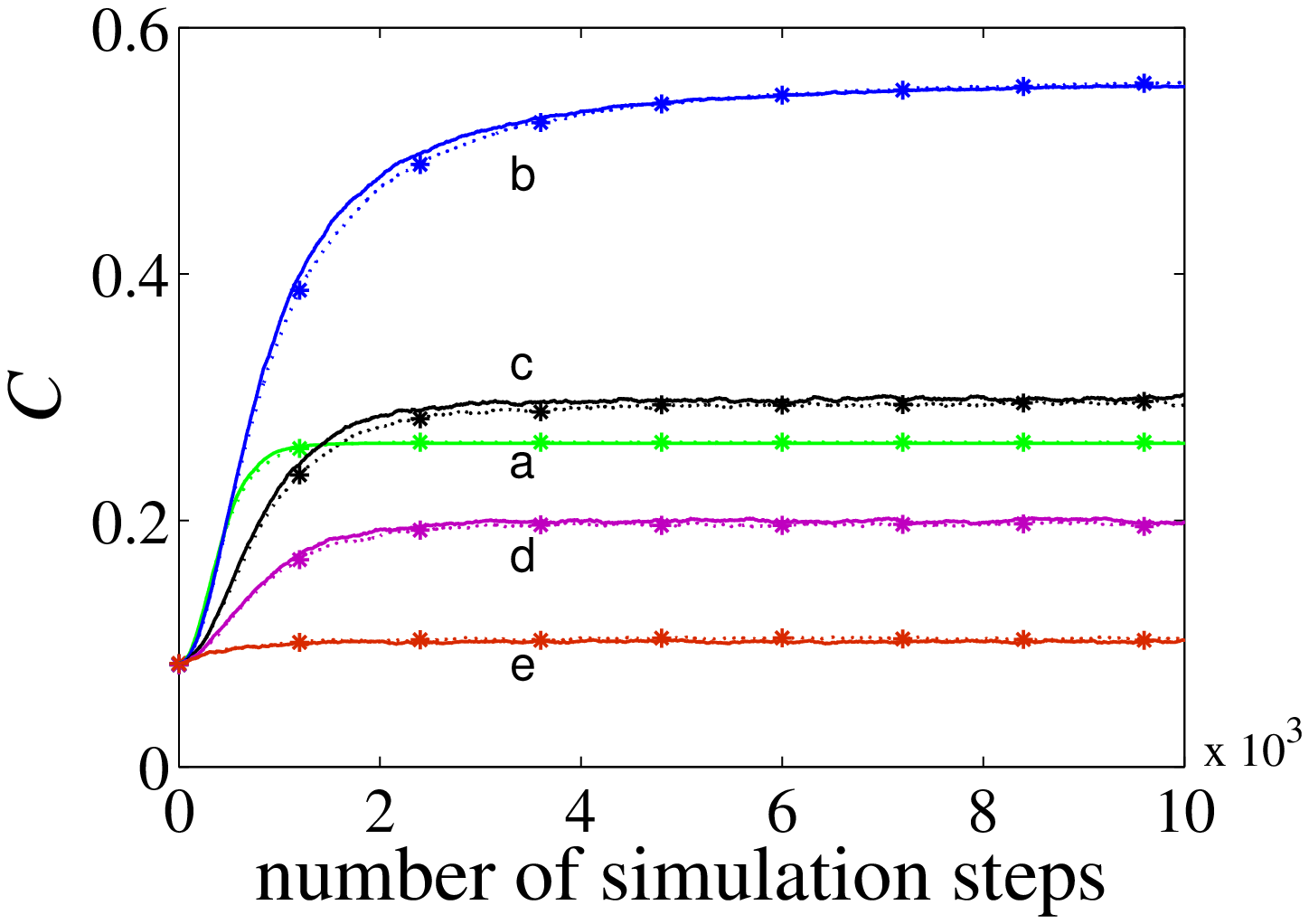}
\includegraphics[scale=0.28]{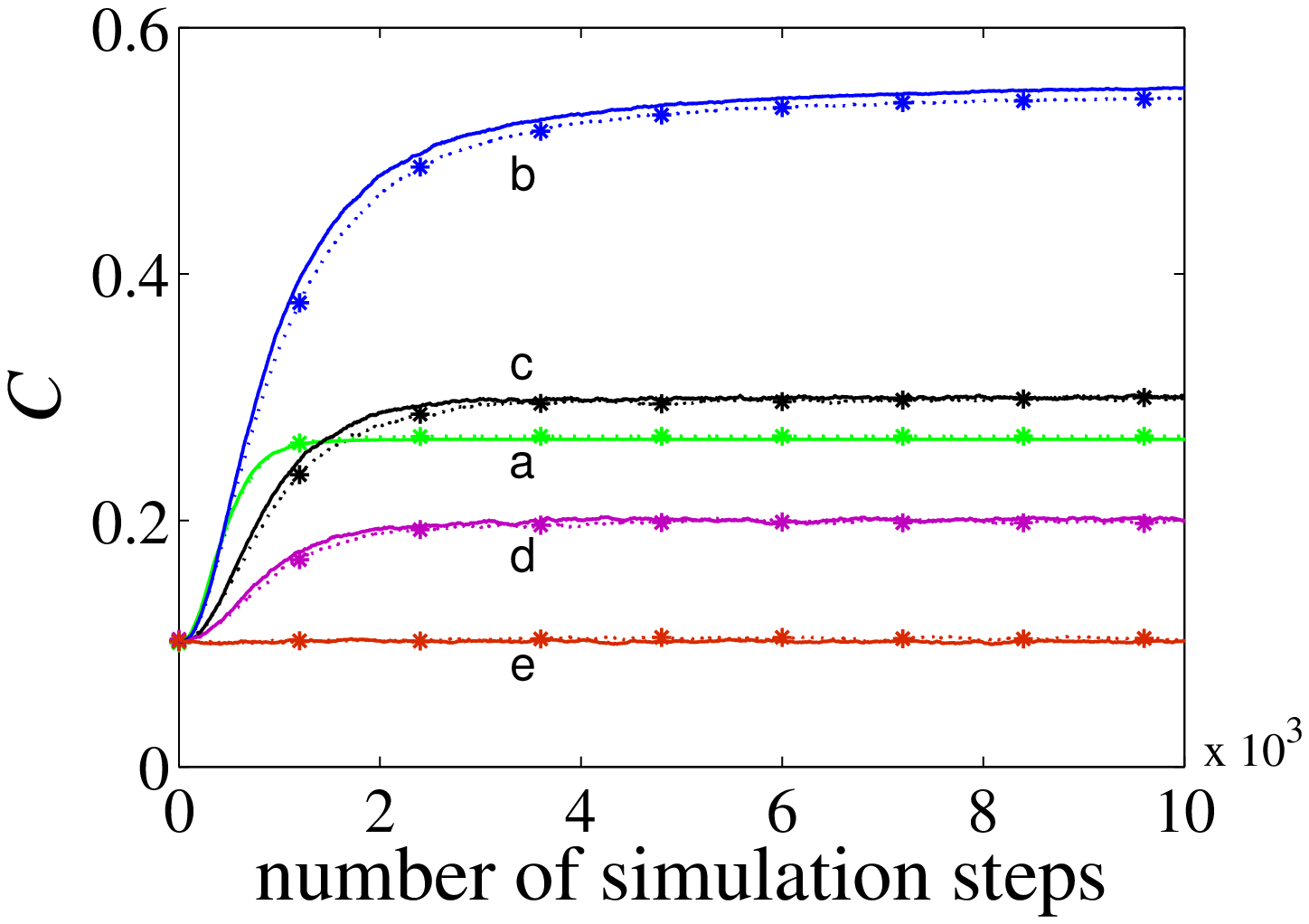}
\caption{\label{fig-10kcluster}Evolution of clustering during rewiring, starting from homogeneous (left) and heterogeneous (right) networks. The plots show the average of 100 simulations with $R = \sqrt{6/\pi}$, $\sqrt{10/\pi}$, $\sqrt{20/\pi}$, $\sqrt{30/\pi}$ and $R$ = $\sqrt{2N}/2$ (green (a), blue (b), black (c), purple (d) and red (e) lines, respectively), where the solid and dotted ($\star$) lines correspond to link- and node-based selection, with $N$ = 100 and $\langle k \rangle$ = 10.}
\end{figure}

Fig.~\ref{fig-10kcluster} shows the evolution of clustering during rewiring for a range of radii $R$, and with both selection methods, as above. As expected, smaller values  of $R$, but such that $\langle k \rangle \ll n$ still holds, lead to higher levels of clustering. However, when $R$ is such that  $\langle k \rangle \gg n$, clustering decreases as rewiring will be limited by the low number of potential targets for rewiring in local areas. This means that many long-range links from the original network will be conserved, and thus clustering is pushed to smaller values. Both selection methods produce similar results in both clustering and preferential mixing for a variety of $R$ values, with both homogeneous and heterogeneous starting networks.

\begin{figure}[h!]
\includegraphics[scale=0.28]{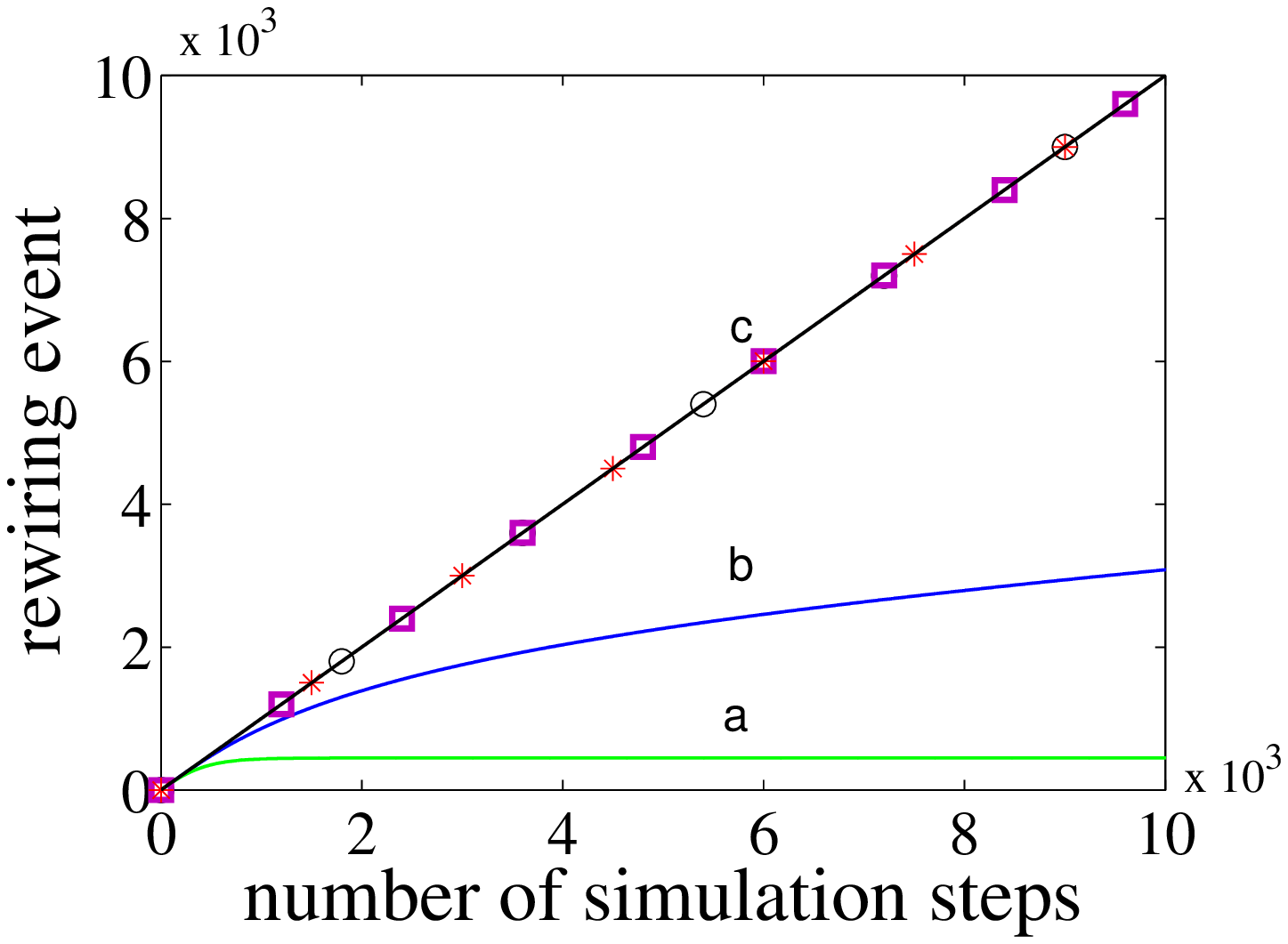}
\includegraphics[scale=0.28]{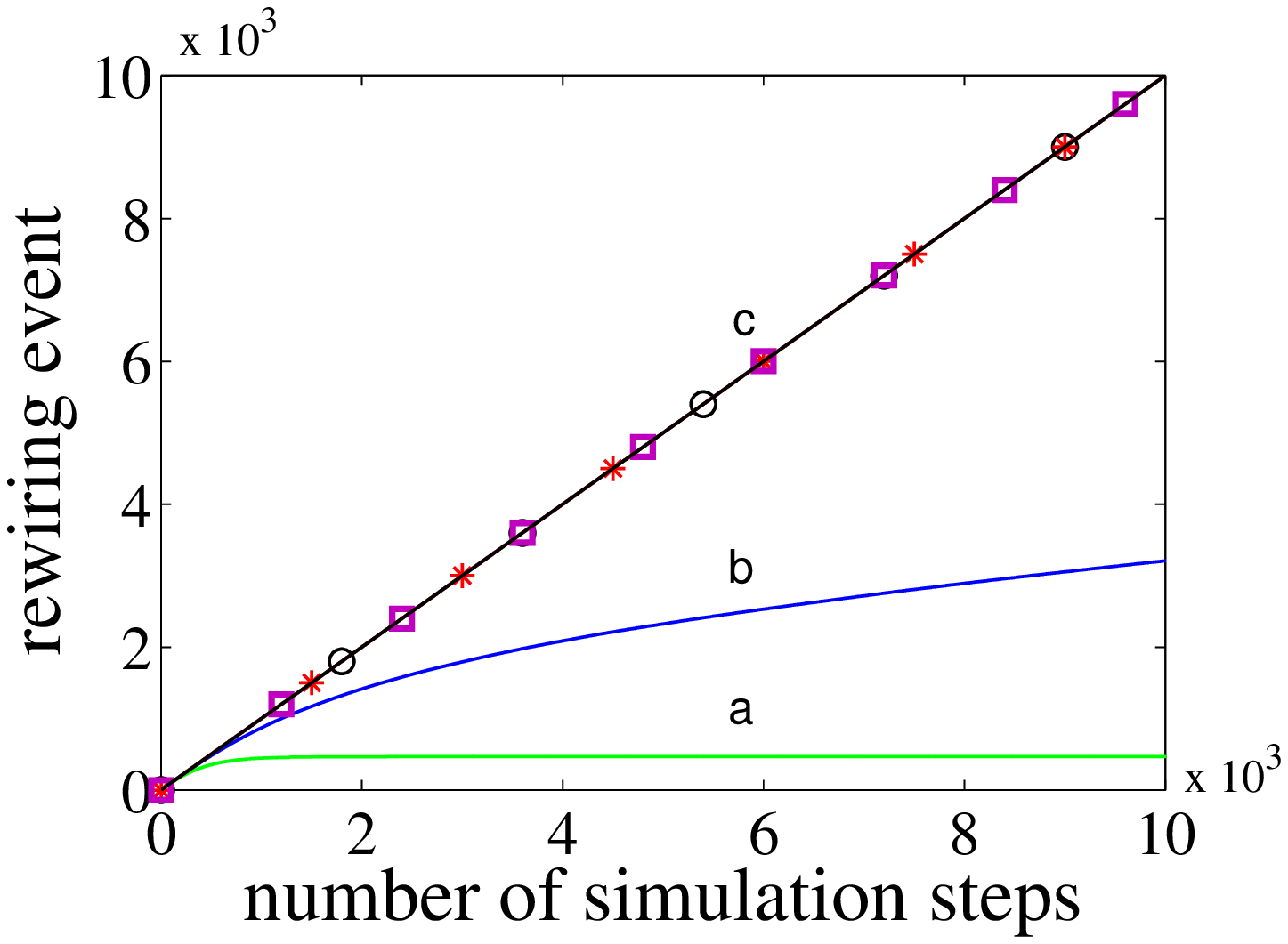}
\caption{\label{fig-10kWevent}Evolution of the rewiring process, starting from homogeneous (left) and heterogeneous (right) networks with node-based selection. The plots show the average of 100 simulations with $R = \sqrt{6/\pi}$, $\sqrt{10/\pi}$, $\sqrt{20/\pi}$, $\sqrt{30/\pi}$ and $R$ = $\sqrt{2N}/2$ (green (a), blue (b), black (c, o), purple (c, $\Box$) and red (c, $\star$) lines, respectively), with $N$ = 100 and $\langle k \rangle$ = 10.}
\end{figure}

It is observed that across all values of radius $R$, given enough time, clustering stabilises. This begs the question of how the rewiring process operates throughout the simulation, especially for large $R$. In Fig.~\ref{fig-10kWevent}, we examine how the number of successful rewiring events depends on the simulation step when using node-based selection for both homogeneous and heterogeneous networks.
As expected, with a small value of $R$, the rewiring process evolves quickly to a stable equilibrium, whereas, for a large value of $R$, it continues throughout the simulation.
Interestingly, for large values of $R$, even when there are still prospective links/nodes to be rewired, clustering of the network is no longer affected (see Fig.~\ref{fig-10kcluster} and Fig.~\ref{fig-10kWevent} where $R = \sqrt{20/\pi}$, $\sqrt{30/\pi}$). Intuitively, this can be explained as follows. Since there are many available target nodes to rewire to in a local area, a node, with say $k$ contacts, proceeds to randomly connect to $k$ nodes within its local area. If the local area is not extremely large, and for relatively dense networks, this process will lead to an initial increase in clustering. Since the area holds more candidates for rewiring than the number of neighbours a node has, link rewiring will continue and other nodes from the same area will be chosen. However,  this will lead to no significant further increase in clustering, except small movements around the equilibrium value.

\subsection{\label{sec:clustering}Computing clustering}
\subsubsection{\label{sec:left-formula}$n \ll \langle k \rangle$: small areas but high degree}
We aim to derive an analytical approximation for clustering by concentrating on the case when, on average, the number of nodes in a circle of radius
$R$ is less than the average degree in the network. In addition, we consider the situation when all possible links have been rewired. Due to having limited options for rewiring locally, we can assume that at the end of the rewiring process almost all of local connections have been realised. We will focus on a typical node and its neighbours within distance $R$ and beyond, noting that two nodes within a circle of radius $R$ are not necessarily at a distance of less than $R$ from each
other.

Let us introduce some notation. Let $B$ be the number of nodes within a radius $R$ from a given node, and not including the node at the centre. $B$ itself is a random variable. Let $k$ be the degree of the node at the centre of the circle ($k$ is therefore also a random variable). To compute the clustering of the central node we seek to establish the number of links between the neighbours of the node. We break this down into links between neighbours who are within the circle, links between internal and
external neighbours and finally links between nodes that are exclusively outside the circle. Counting multiplicatively, the total number of possible triangles is:
$$B(B-1)+2B(k-B)+(k-B)(k-B-1)=k(k-1).$$

We now set out to find the probability of connections existing between the three different types of edges. First, we work out the probability of two interior nodes being connected. This can be done by considering a circle of radius $R$ and then an arbitrary point within it. The probability that the second node will be within distance $R$ from the initial node will be proportional to the overlap area, $A_{overlap}$, between the original circle and the circle of radius $R$ centred around the first random point. Hence, the probability that the distance between the two random points within the circle is less then $R$ is simply:
$$P(d<R)=\frac{A_{overlap}}{\pi R^2}.$$

To determine $A_{overlap}$, we first work out the density function for the distance of the first point from the centre. However, when placing nodes at random in a circle, the uniform random number has to be scaled with the $\sqrt{\cdot}$ function. Effectively, a good or valid random choice for the distance from the centre is not
$unifran(0,1)R$, but $\sqrt{unifran(0,1)}R$. This means that the density function for the distance from the centre of a randomly and uniformly placed node is: $\rho(r)=\frac{2r}{R^2}$. This integrates to 1 for $r$ going from $0$ to $1$. Knowing the distance $r$ between the two points, we average the well known area for the intersection of two circle or radii $R$ and with distance $r$ between their centres, that is:
$$A_{overlap}(r, R)=2R^2\cos^{-1}(\frac{r}{2R})-\frac{1}{2}r\sqrt{4R^2-r^2}.$$

Hence, the probability that two nodes within a circle of radius $R$ are less than $R$ apart is given by:
$$q=\int_{0}^{1}A_{overlap}(r,R)\frac{2r}{R^2}dr,$$
and the number of triangles that are forming between interior
nodes is $B(B-1)q$.

We now focus on the probability of links existing between the remaining non-connected interior-interior nodes (of which there are $B(B-1)(1-q)$), as well as between interior-exterior (i.e. $2B(k-B)$) and exterior-exterior (i.e. $(k-B)(k-B-1)$) nodes. In general, we can state that if the distance between two nodes is less than $R$ then at the end of the simulation they will have formed a link. The probability that the distance between two randomly placed nodes is less than $R$ is the ratio between the area of the circle/local area with respect to the total area. Thus, with probability $\frac{\pi R^2}{N}$, two nodes are less than $R$ apart and are connected with probability 1. With probability $1-\frac{\pi R^2}{N}$, these nodes will be more than $R$ away and therefore will be connected by the long-range links that remain at the end of the rewiring process. However, the average number of such links is $(k-B)N$ with short-range links accounting for $BN$. Thus assuming that long-range links are distributed at random across all possible long-range pairs we get that the probability of such a link existing is
$$p_{lr}=\frac{(k-B)N}{N(N-1)(1-\frac{\pi R^2}{N})}.$$
Hence, a random pair of nodes forms a link with probability
$$\frac{\pi R^2}{N}+(1-\frac{\pi R^2}{N})p_{lr}=\frac{k+1}{N-1}-\frac{B+1}{N(N-1)} \sim \frac{k+1}{N-1},$$
since $\frac{B+1}{N(N-1)}$ is likely to be small. However, surprisingly, this value is very close to what is the initial probability of a link existing when the network is connected up according to the Erd\H{o}s R\'enyi model. In this case, the probability of a link existing is $\frac{k}{N-1}$ which is also the measure of clustering for the initial network since all links are placed at random and thus where a node has two neighbours, the probability of them being connected is $C=\frac{k}{N-1}$. However, at the end of the rewiring process we get that clustering should be well approximated by

\begin{eqnarray}\label{c-left}
          C_L &=&\frac{B(B-1)q+p_{lr}B(B-1)(1-q)}{k(k-1)} \nonumber \\
            & &+\frac{(\frac{k+1}{N-1}-\frac{B+1}{N(N-1)}) \left[2B(k-B)\right]}{k(k-1)} \nonumber \\
            & &+\frac{(\frac{k+1}{N-1}-\frac{B+1}{N(N-1)}) \left[(k-B)(k-B-1)\right]}{k(k-1)}.
\end{eqnarray}

We expect that when clustering is high, the $B(B-1)q$ term dominates. We can also suggest a simpler formula for $C$, namely, one that assumes that almost all interior neighbours of a central node will become connected and the contribution from other pair types towards clustering is small. On the one hand, this overestimates clustering when looking at connections between interior nodes, as these could be apart by more than distance $R$. On the other hand, it underestimates clustering as some interior-exterior and
exterior-exterior nodes could still be connected. This formula gives
$$C_{a}=\frac{B(B-1)}{k(k-1)}.$$

Both formulas above work on average or expected values. As noted previously, $k$ and $B$ can be treated as random variable with some distribution. An analytic or semi-analytic expression for these would make it possible to numerically evaluate our two approximations and compare them to clustering measured from simulations.

\subsubsection{\label{sec:right-formula}$n \gg \langle k \rangle$: large areas but low degree}
Let us use the same definition of $B$ and $k$ as in the previous section, but here $B > k$. Our analysis will focus on a typical node out of the $k$ nodes in the area. Since the probability of two nodes within a circle of radius $R$ being connected is $q$ and there are $B-1$ nodes in total available to form links, clustering should be approximated by
\begin{equation}
    C_R = q\frac{k-1}{B-1}.
\label{c-right}
\end{equation}
This formula works on the assumption that the centre node forms triangles only within its local area since $B > k$.

\begin{figure}[h!]
    \includegraphics[scale=0.28]{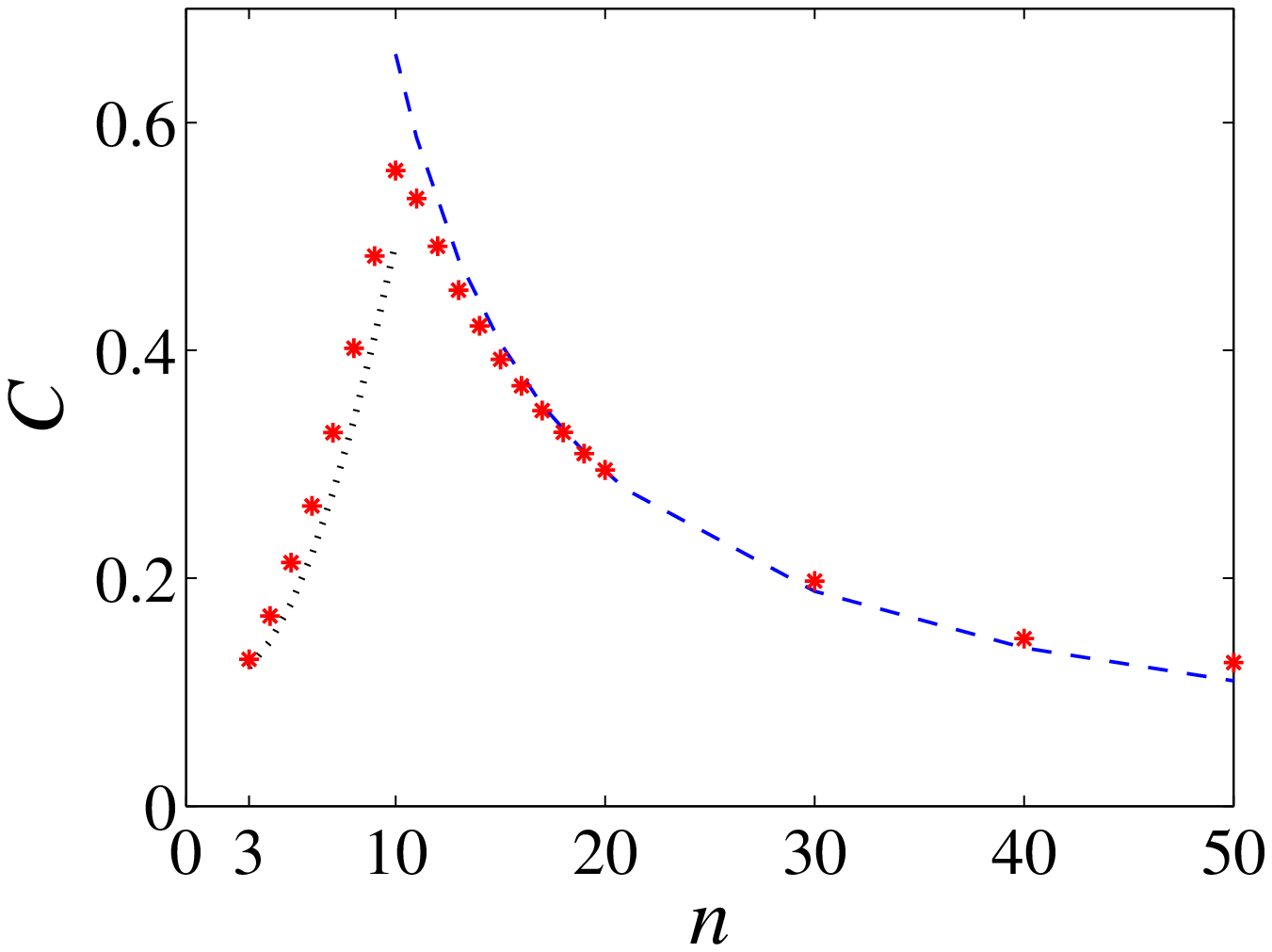}
    \includegraphics[scale=0.28]{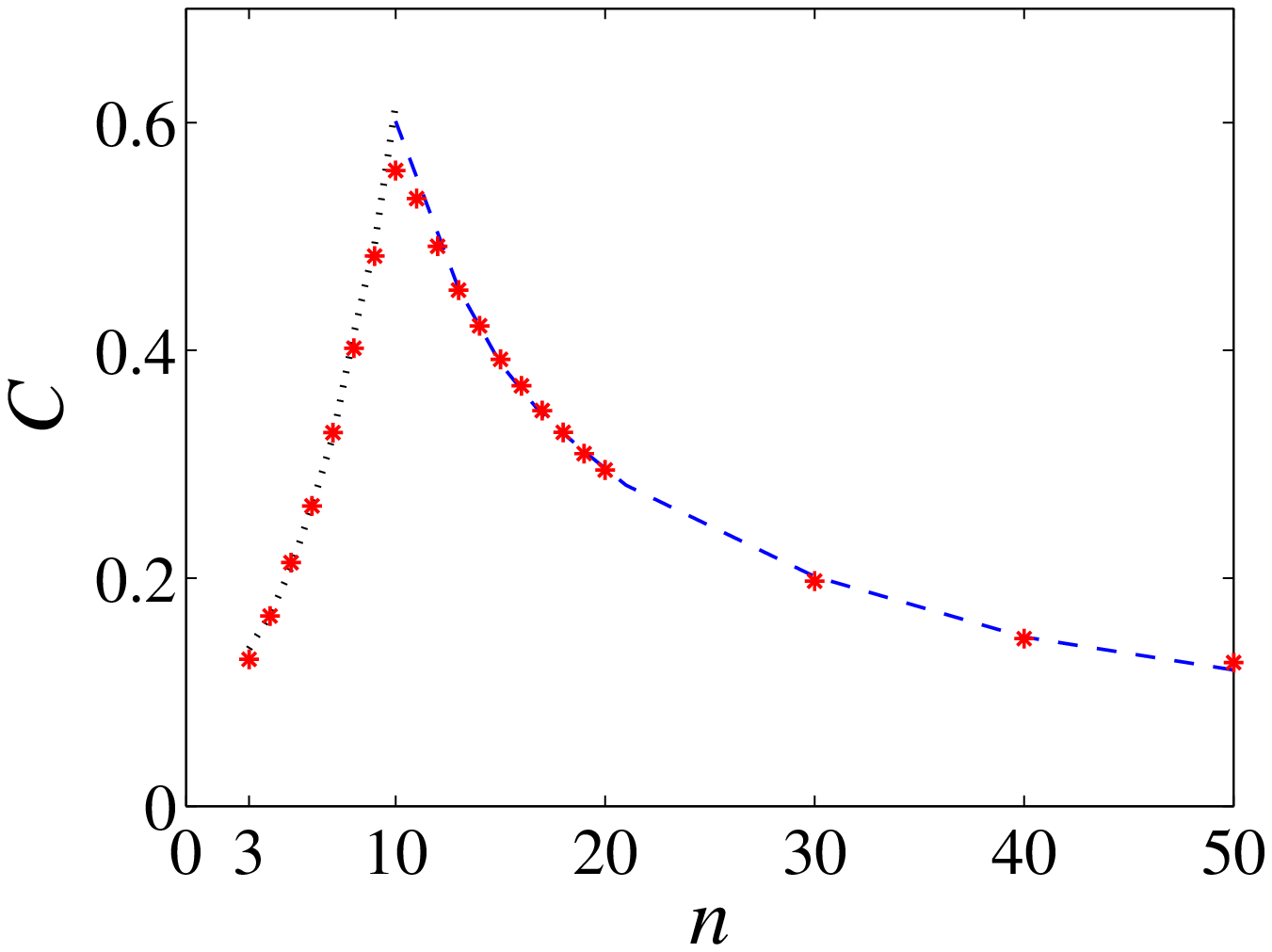}
    \caption{\label{fig-clust-formulas}Clustering at the end of simulations starting from homogeneous networks with node-based selection.
     Simulation results (red $\star$) are compared with analytic formulas (Eq.~(\ref{c-left}) (black dotted) and Eq.~(\ref{c-right}) (blue dashed)), with $k=\langle k \rangle$.
     In the left panel we use the formulas with average $(B,k)$ values. In the right panel we use $(B,k)$'s joint distribution computed from simulation. The plots show the     average of 100 simulations with $N$ = 100 and $\langle k \rangle$ = 10.}
\end{figure}

Both Eq.~(\ref{c-left}) and Eq.~(\ref{c-right}) are shown in Fig.~\ref{fig-clust-formulas}. Here, we present only the case of homogeneous networks with node-based selection, due to the two rewiring methods giving very similar clustering values, see Fig.~\ref{fig-10kcluster}. The left panel of Fig.~\ref{fig-clust-formulas} uses average $(B,k)$ values such that all centre nodes have $B$ nodes within a radius $R$ and have degree $k$. However, this is an approximation since in reality $B$ and $k$ are random parameters and have a joint distribution. When accounting for this heterogeneity by computing the joint parameter distribution from simulation, the agreement significantly improves as shown in the right panel of Fig.~\ref{fig-clust-formulas}. Here, we randomly choose 5\% of $N$ nodes to be centre nodes and count the
true values of $B$ and $k$ to compute the clustering.

While the analytic formulas for the clustering values are derived for the limiting cases of $n \ll \langle k \rangle$ and $n \gg \langle k \rangle$, a close examination of
Fig.~\ref{fig-clust-formulas} reveals that agreement with simulation is maintained close to the $n \simeq \langle k \rangle$ regime. Moreover, the same figure shows that
the maximum value of clustering is achieved for $n \simeq \langle k \rangle$. By using this value in the analytic formulas, i.e., $B=n-1=\langle k \rangle -1$, and by neglecting the small terms leads to
\begin{equation*}
    C_L = q\left(1-\frac{2}{\langle k \rangle}\right)\,\,\, \text{and} \,\,\, C_R=q\left(1+\frac{1}{\langle k \rangle-2}\right),
\end{equation*}
which shows that clustering will be dominated by the probability, $q$, that two nodes within a circle of radius $R$ are less than a distance $R$ apart. The value of $q$ is independent of $R$ and it is $q \sim 0.58$, as confirmed by our figure. While, $C_L$ underestimates and $C_R$ overestimates clustering at $n=\langle k \rangle$, it is worth noting that using $n=\langle k \rangle+1$ or $B=\langle k \rangle$ in both formulas, i.e., Eq.~(\ref{c-left}) and Eq.~(\ref{c-right}), we get
\begin{equation*}
    C_L = C_R=q.
\end{equation*}
Hence, we can conclude that clustering can be maximised if the expected number of nodes in the local area is very close or identical to the expected degree of a node.
Such a setup will ensure that all potential neighbours can be drawn from inside a local area and clustering will be dominated by the probability, $q$, that two nodes within a circle of radius $R$ are less than a distance $R$ apart.

For large $n$, $n \rightarrow N$, the reasoning that lead to working out $q$ breaks down, since for large $R$ values almost all nodes are in the same unique area. This effectively means that $q \rightarrow 1$ and thus $C_R \rightarrow \frac{\langle k \rangle-1}{N-2}\simeq \frac{\langle k \rangle-1}{N-1}$ (for large $N$), which is the value of clustering in a random network.

\begin{figure}[h!]
\begin{center}
    \includegraphics[scale=0.28]{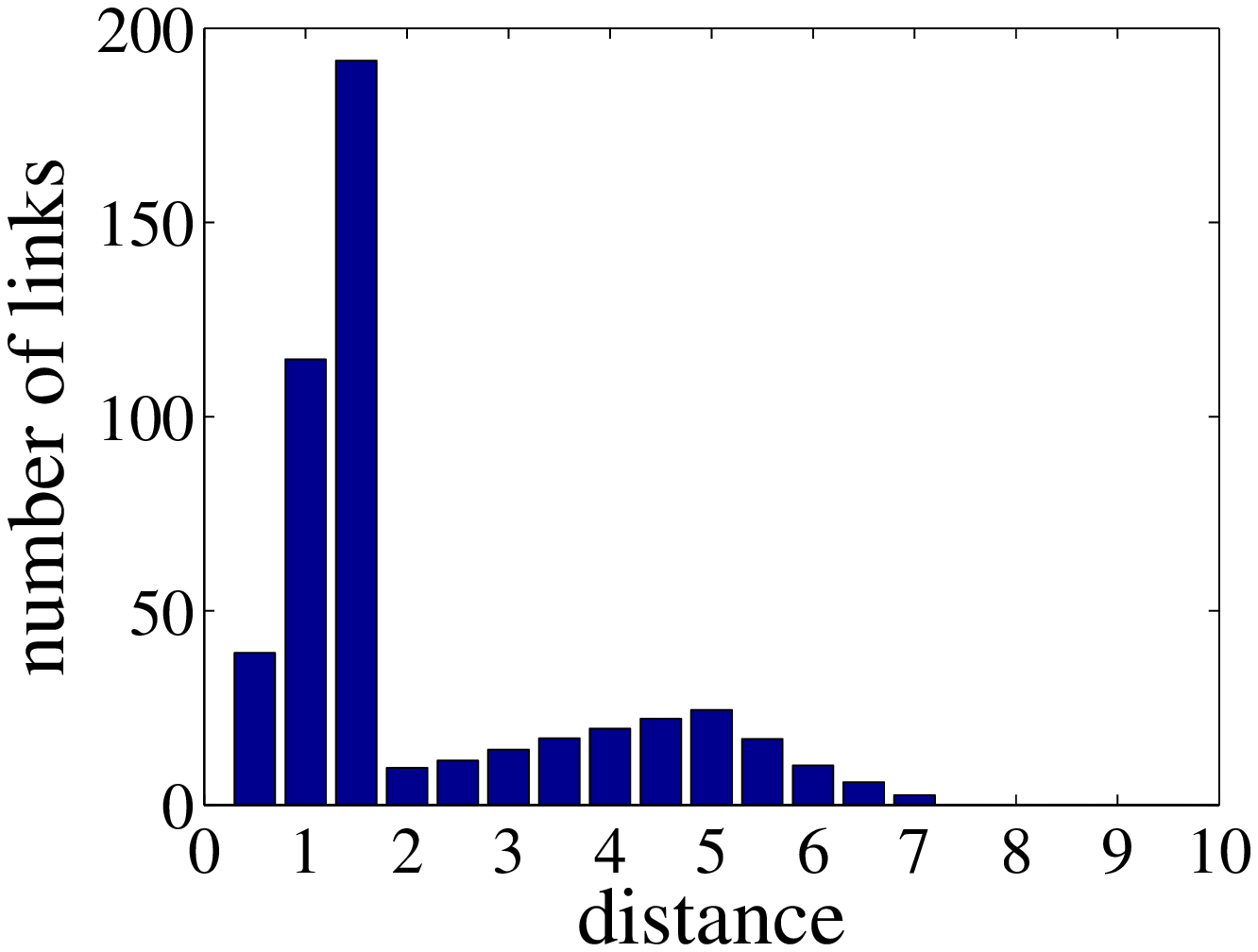}
    \includegraphics[scale=0.28]{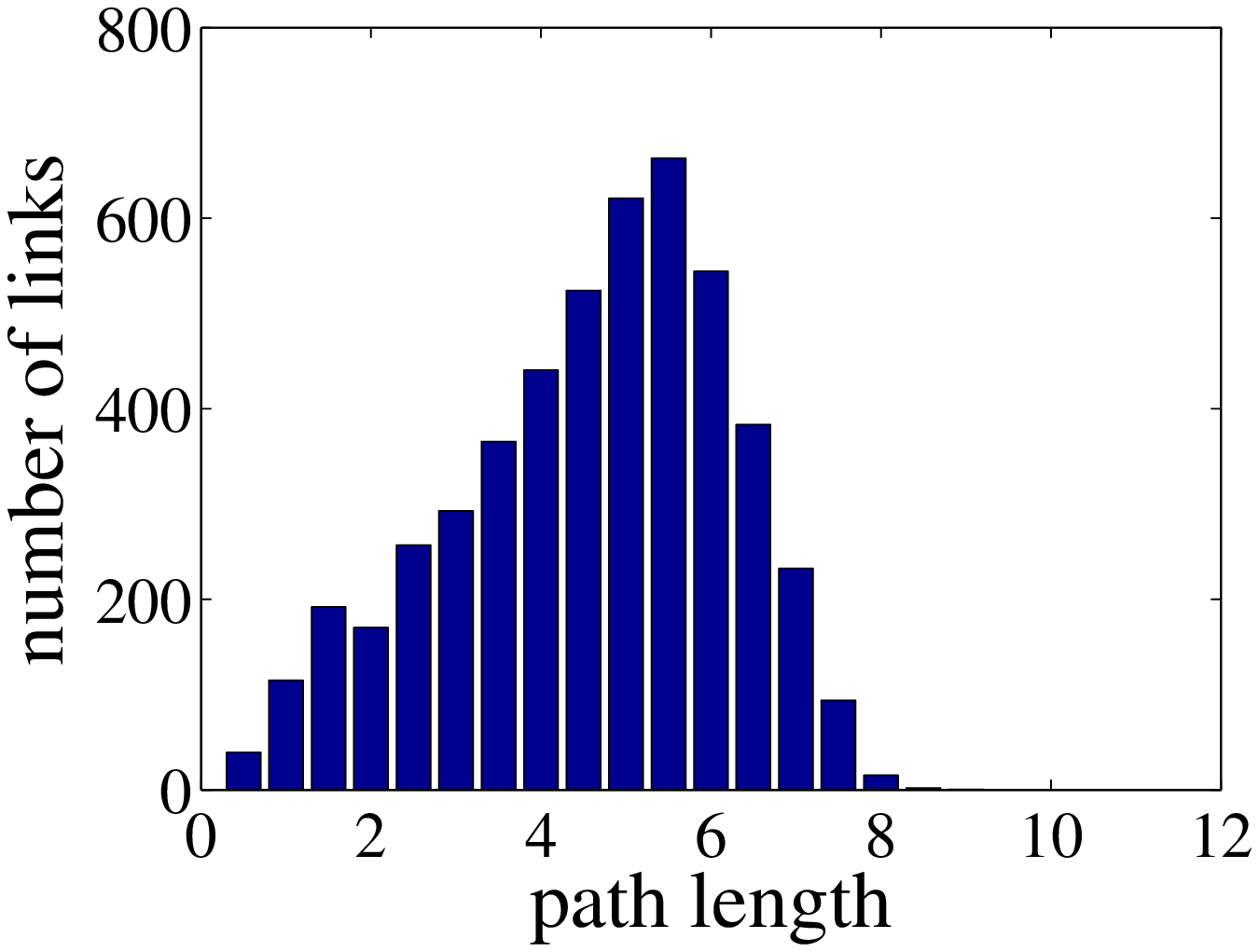}
    \includegraphics[scale=0.28]{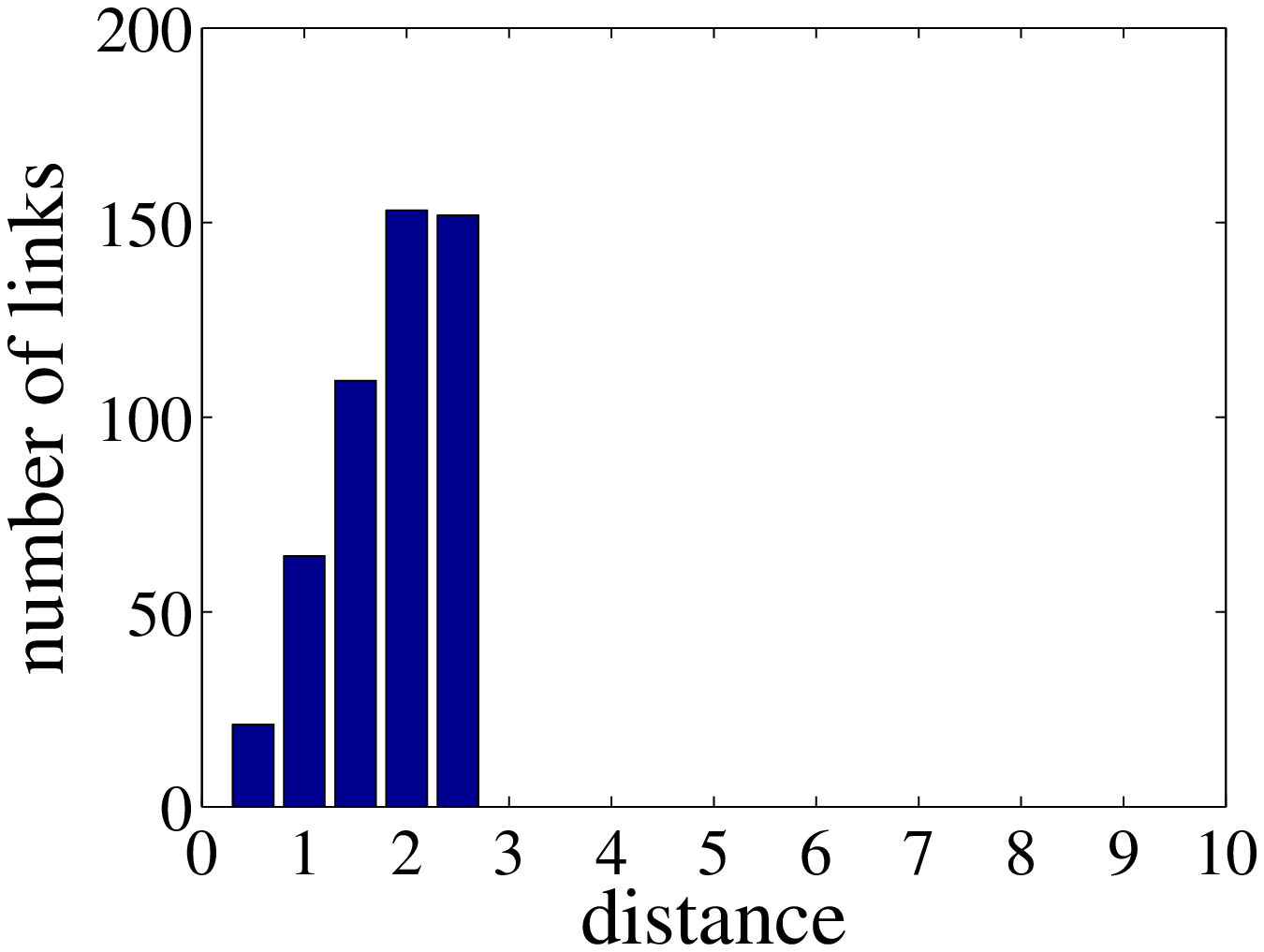}
    \includegraphics[scale=0.28]{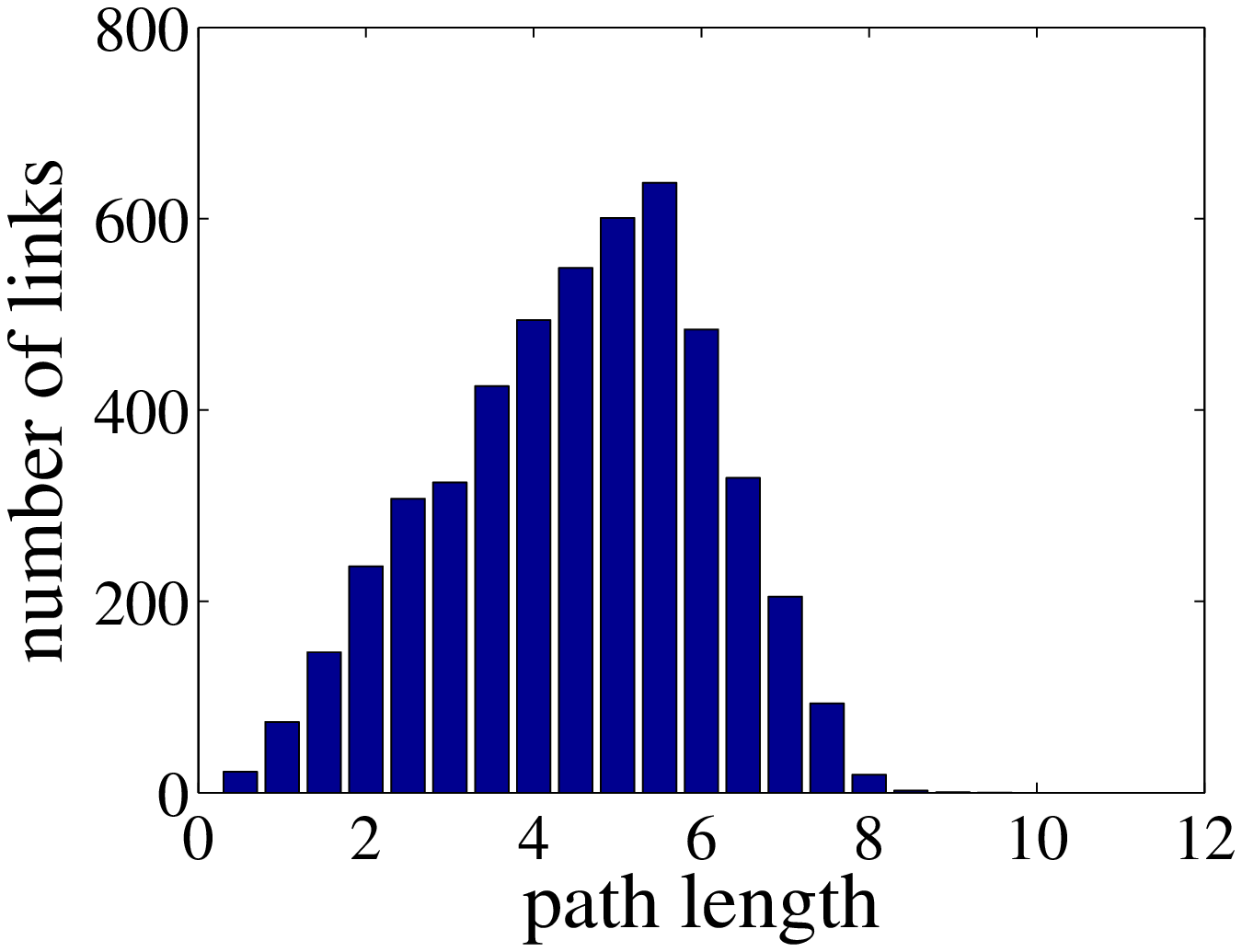}
    \caption{\label{fig-dis-DL}Distribution of distance between $i$ and $j$ if $g(i,j)=1$ and distribution of path length at the end of simulations starting from homogeneous networks with node-based selection. The plots show the average of 100 simulations for $n = 7$ (top) and $ n= 18$ (bottom) with $N$ = 100 and $\langle k \rangle$ = 10.}
\end{center}
\end{figure}

\begin{figure}[h!]
\begin{center}
    \includegraphics[scale=0.35]{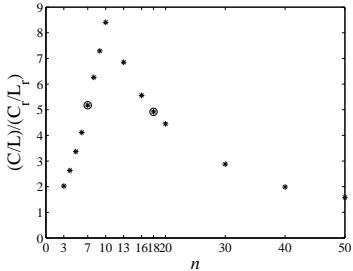}
    \caption{\label{fig-CL-CrLr}The small-worldness index $(C/L)/(C_r/L_r)$ at the end of simulations starting from homogeneous networks with node-based selection. The plots show the average of 100 simulations with $N$ = 100 and $\langle k \rangle$ = 10.}
\end{center}
\end{figure}

From Fig.~\ref{fig-clust-formulas}, we note that networks with the same level of clustering can be generated with both $n \ll \langle k \rangle$ and $n \gg \langle k \rangle$. This begs the interesting question of whether structural differences exist in these networks. We examined a number of network characteristics, including path length distribution and distribution of true link lengths. Fig.~\ref{fig-dis-DL} shows the distribution of distance for all links as well as the distribution of path length, for $n = 7$ and $n = 18$. As expected, with a large value of $n$, the rewiring will be able to rewire all links. Thus, the final network has all its links with length less than or equal to the value of $R$ (see distribution of distance in Fig.~\ref{fig-dis-DL} when $n = 18$). The final networks show a slight difference in mean path length, $L(n = 7)\approx4.33$ and $L(n=18)\approx4.26$, even though their distributions of distance are significantly different. To further highlight the different network structures, Fig.~\ref{fig-CL-CrLr} shows the small-worldness index of each final network as a function of $n$. This index is obtained by computing the ratio of $C/L$ divided by the ratio of $C_r/L_r$ where $C_r$ and $L_r$ are the clustering and mean path length respectively of the equivalent randomised network.

\subsection{\label{sec:labelling}Rewiring within local areas with $SI$ labelling}
To get closer to the full model (i.e., coupled epidemic dynamics and rewiring) and to gain more insights into the properties of the adaptive network, we now consider the scenario in which each node is assigned a disease status. Using the analogy of simple epidemic models, such as the $SIS$ model, nodes are labelled at random as susceptible, $S$ nodes, with probability $p_s$ and infected, $I$ nodes, with probability $p_i=1-p_s$. We consider the network when the rewiring mechanism makes use of node labels, but without the full epidemic dynamics. This means that whilst the numbers of $S$ and $I$ are constant, the number of each type of links changes depending on type; namely, the number of $SI$ decreases, the number of $SS$ links increases and the number of $II$ remains constant, thus changing the structure of the network. Provided that $S_0=p_sN$ and $I_0=(1-p_s)N$, the initial link counts for $SS$, $II$ and $SI$ links are $S^2_0 \langle k \rangle / 2N $, $I^2_0 \langle k \rangle / 2N $ and $S_0 I_0 \langle k \rangle / N$, respectively, where each link is uniquely counted. When one of the $SI$ links is cut and a new $SS$ link is formed, it is obvious that the total number of $SS$ links increases relative to the (decreasing) number of $SI$ links, and therefore, most $S$ nodes in the network will evolve higher degrees.

This adaptive rewiring rule can lead to the network dividing into two sub-networks: one containing only $S$ nodes and $SS$ connections, and the other $I$ nodes with $II$ connections. Of course, this is not unique to the introduction of local rewiring constraints, i.e., $R<\sqrt{2N} / 2$.
Further, it should be noted that it is possible that not all $SI$ links are cut. This can happen when there is a very small number of $S$ nodes compared to a large number of $I$ nodes or when the local neighbourhood or radius is very small. In this case, not all $SI$ links can be cut since reconnection would lead to multiple links, which we do not allow.

To simplify the dynamics of the adaptive network, we start with $S_0=80\% $ of $N$ and $I_0=20\%$ of $N$, and we allocate node labels at random. As previously, an $SI$ link is chosen at random, and the $S$ node within this link reconnects to another $S$ node in its local area, provided that such node exists. Otherwise, the rewiring step is abandoned and a new $SI$ link is selected. The simulation or rewiring completes when either all $SI$ links have been rewired or the remaining links cannot be rewired due to a lack of available $S$ nodes in the local areas.\\

\noindent \textbf{Impact of rewiring on the degree distribution of the network}
\\
To explore the impact of the rewiring dynamics (whereby only $SS$ links can be formed) on network degree, we consider changes in degree distribution when starting with either homogeneous or heterogeneous networks.

\noindent (a) \textit{Heterogeneous networks:}

When starting from a heterogeneous network at time $t=0$, the network has a degree distribution given by the binomial distribution, namely $p(k)={N-1 \choose k} p^k (1-p)^{N-1-k}$, where $p=\langle k \rangle /(N-1)$, and the average degree of both susceptible and infected nodes is equal to $\langle k \rangle$.
We assume that the degree distribution of $S$ and $I$ nodes remain random throughout the simulation, and is binomial. First, let us consider the degree distribution of $S$ nodes. We start by calculating the average degree of $S$ nodes at time $t$. Let us define $\Delta k_S(t)$ as the rate of change of the average degree of $S$ nodes, and assume that $\Delta k_S(t)$ depends on the number of $SI$ links that are being cut at time $t$. Since the average degree of $S$ nodes at the end of the simulations (when all $SI$ links have been cut) is given by $(1+i_0) \langle k \rangle$~\cite{Gross2006}, where $i_0=I_0/N$, $\Delta k_S(t)$ can be computed as
\[
    \Delta k_S(t) = \Big[(1+i_0)\langle k \rangle - \langle k \rangle \Big] \frac{[SI]^{cut}(t)}{[SI]_0} = i_0 \langle k \rangle \frac{[SI]^{cut}(t)}{[SI]_0},
\]
where $[SI]_0$ is the initial number of $SI$ links and $[SI]^{cut}(t)$ is the total number of $SI$ links that have been cut upto time $t$. Then, as we know that all $S$ nodes have degree $\langle k \rangle$ at $t=0$, and the degree can only increase by  $\Delta k_S$ due to the rewiring process, we can calculate the average degree of a $S$ node as
\begin{eqnarray*}
\langle k_S \rangle (t) &=& \langle k \rangle + \Delta k_S(t) \\
                        &=& \langle k \rangle + i_0 \langle k \rangle \frac{[SI]^{cut}(t)}{[SI]_0} \\
                        &=& \Big[ 1 + i_0 \frac{[SI]^{cut}(t)}{[SI]_0} \Big] \langle k \rangle.
\end{eqnarray*}
Therefore, the degree distribution of a susceptible node can be written as
\begin{equation}
        P( S = a)_t = {N-1 \choose a} {p^a_{S}}(1-p_S)^{N-1-a},
\label{hete-dis-S}
\end{equation}
where $a=0,1,2,...,N-1$ and $p_S=\frac{\langle k_S \rangle (t)}{N-1}$.

We can use the same methodology to derive $\Delta k_I(t)$, the average degree and the degree distribution of $I$ nodes. However, the degree of $I$ nodes can only decrease by $\Delta k_I$ and using the average degree of $I$ nodes, $i_0 \langle k \rangle $, when all $SI$ links have been cut~\cite{Gross2006}, we get

\begin{eqnarray*}
\langle k_I \rangle (t) &=& \langle k \rangle - \Delta k_I(t) \\
                        &=& \langle k \rangle -\Big[ \langle k \rangle - i_0 \langle k \rangle \Big] \frac{[SI]^{cut}(t)}{[SI]_0} \\
                        &=& \langle k \rangle - s_0 \langle k \rangle \frac{[SI]^{cut}(t)}{[SI]_0} \\
                        &=& \Big[ 1 - s_0 \frac{[SI]^{cut}(t)}{[SI]_0} \Big] \langle k \rangle.
\end{eqnarray*}
Therefore, the degree distribution of an infected node can be written as
\begin{equation}
    P( I = a )_{t} = {N-1 \choose a} {p^a_{I}} (1-p_I)^{N-1-a},
\label{hete-dis-I}
\end{equation}
where  $a=0,1,2,...,N-1$ and $p_I=\frac{\langle k_I \rangle (t)}{N-1}$.
\\

\noindent (b) \textit{Homogeneous networks:}

We now focus on homogeneous networks for which the degree distribution of the network at time $t=0$ is $p(k)= 1$, and the average degree of both susceptible and infected nodes is equal to $k$. Since we apply a random rewiring process, we assume that the network will evolve towards a random network with a binomial distribution, both for $S$ and $I$ nodes. As before, we assume that the average degree of $S$ nodes increases by $\Delta k_S$, and the average degree of $I$ nodes decreases by $\Delta k_I$, which depends on how many $SI$ links are cut. In the case of $S$ nodes, all $S$ nodes start with exactly $k$ links and their degree will increase to $k+1$, $k+2$, $k+3$, ..., $k+S_0-1$. Similarly, all $I$ nodes start with $k$ links and their degree will be decreased to $k-1$, $k-2$, $k-3$, ..., 0. So we have
\[
    \Delta k_S (t) = i_0 {k} \frac{[SI]^{cut}(t)}{[SI]_0},
\]
and the degree distribution of a susceptible node can be written as
\begin{equation}
    P( S = a)_t = {S_0-1 \choose a} {p^a_{S}}(1-p_S)^{S_0-1-a}, 
    \label{homo-dis-S}
\end{equation}
where $a = 0,1,2,...,S_0-1$, $\langle k_S\rangle (0) = k$, $\langle k_S \rangle (t) = \Delta k_S (t)$ and $p_S=\frac{\langle k_S \rangle (t)}{S_0-1}$.

In the case of $I$ nodes, using the same approach as for heterogeneous networks yields
\[
    \langle k_I \rangle (t) = k - \Delta k_I (t) = \Big[ 1 - s_0 \frac{[SI]^{cut}(t)}{[SI]_0} \Big] k ,
\]
and, therefore, the degree distribution of an infected node can be written as
\begin{equation}
    P(I = a)_t = {k \choose a} {p^a_{I}} (1-p_I)^{k-k_I}, a = 0,1,2,...,k,
\label{homo-dis-I}
\end{equation}
where $p_I=\frac{\langle k_I \rangle (t)}{k}$.

\begin{figure}[h!]
\begin{center}
    \includegraphics[scale=0.28]{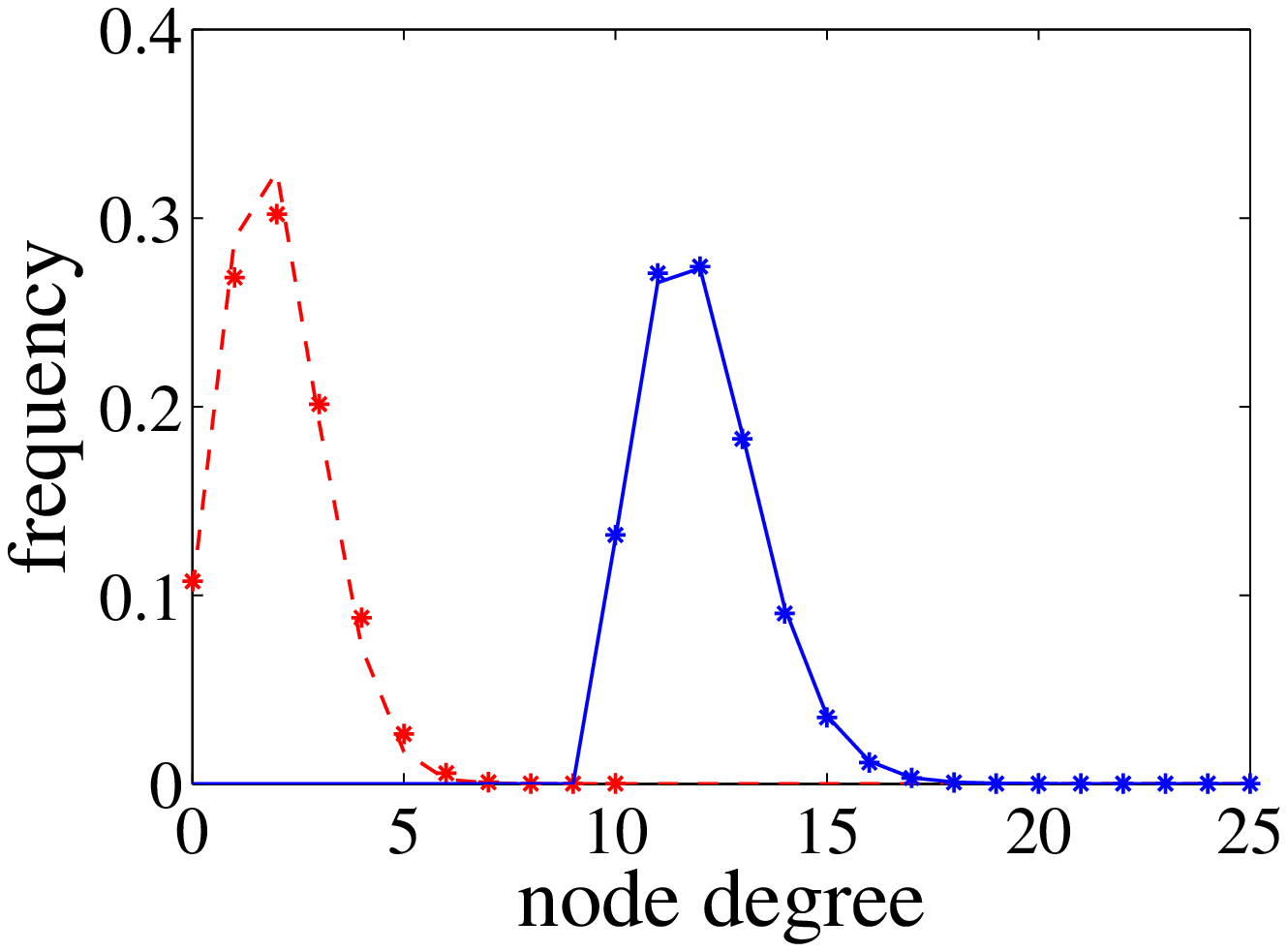}
    \includegraphics[scale=0.28]{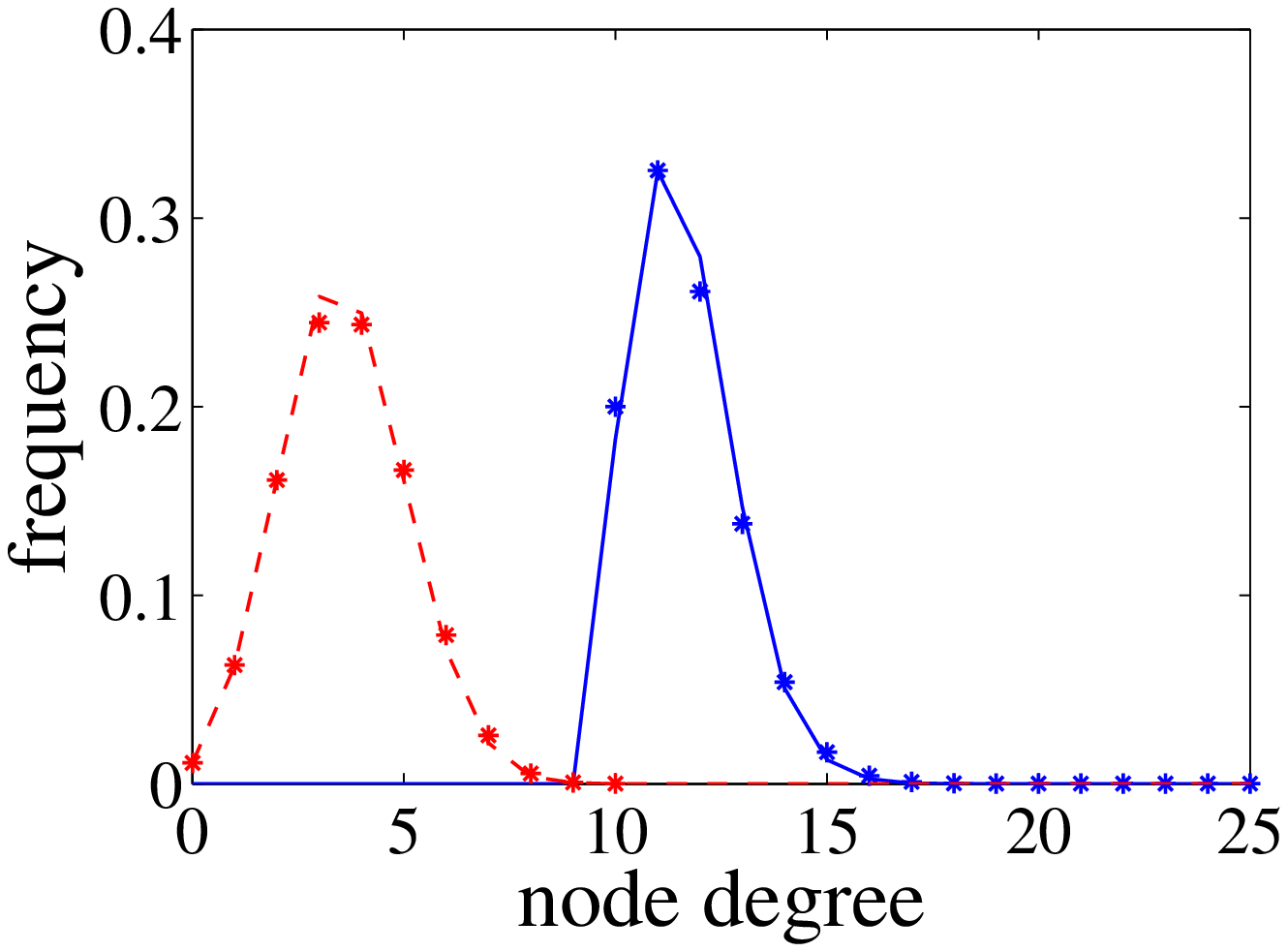}
    \includegraphics[scale=0.28]{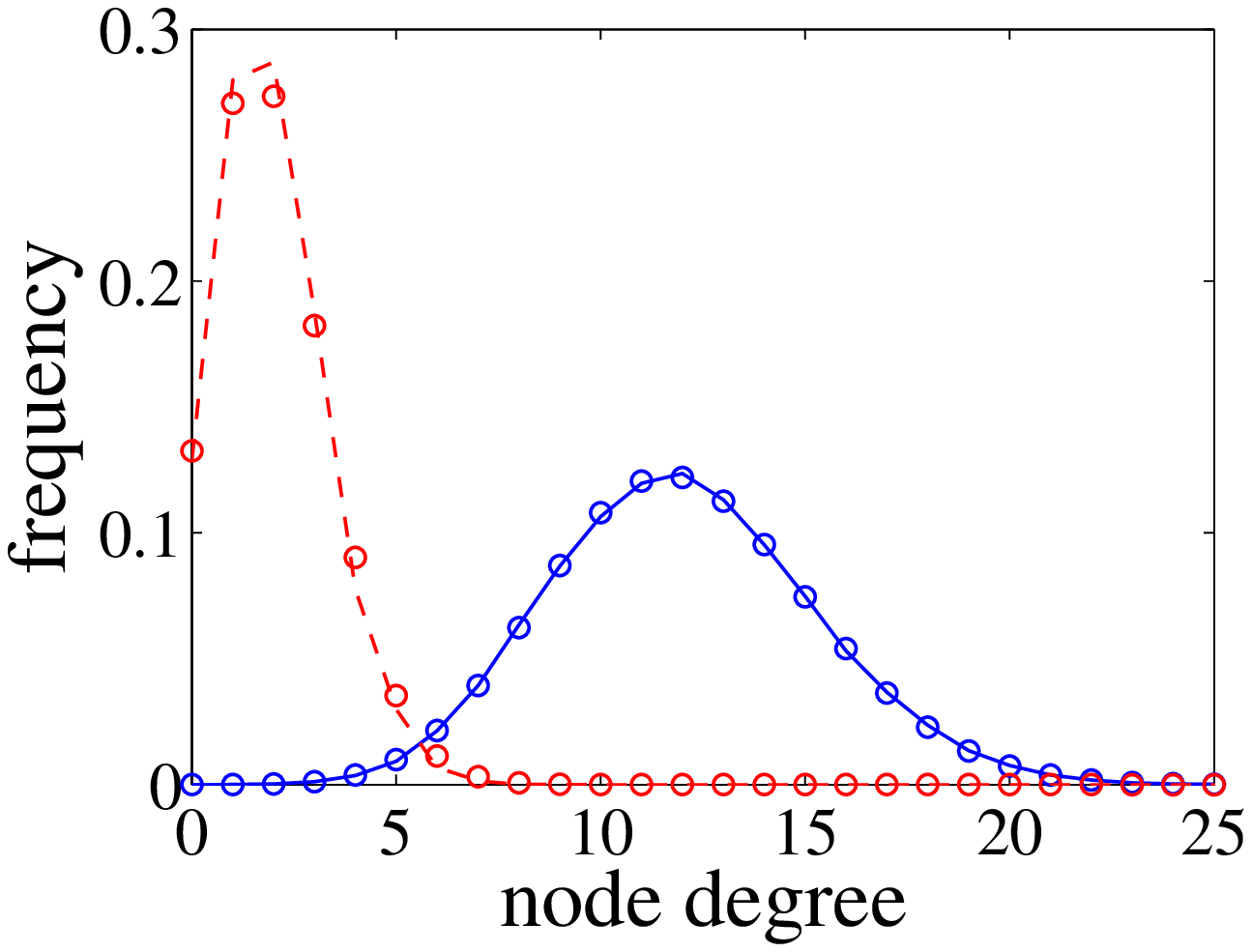}
    \includegraphics[scale=0.28]{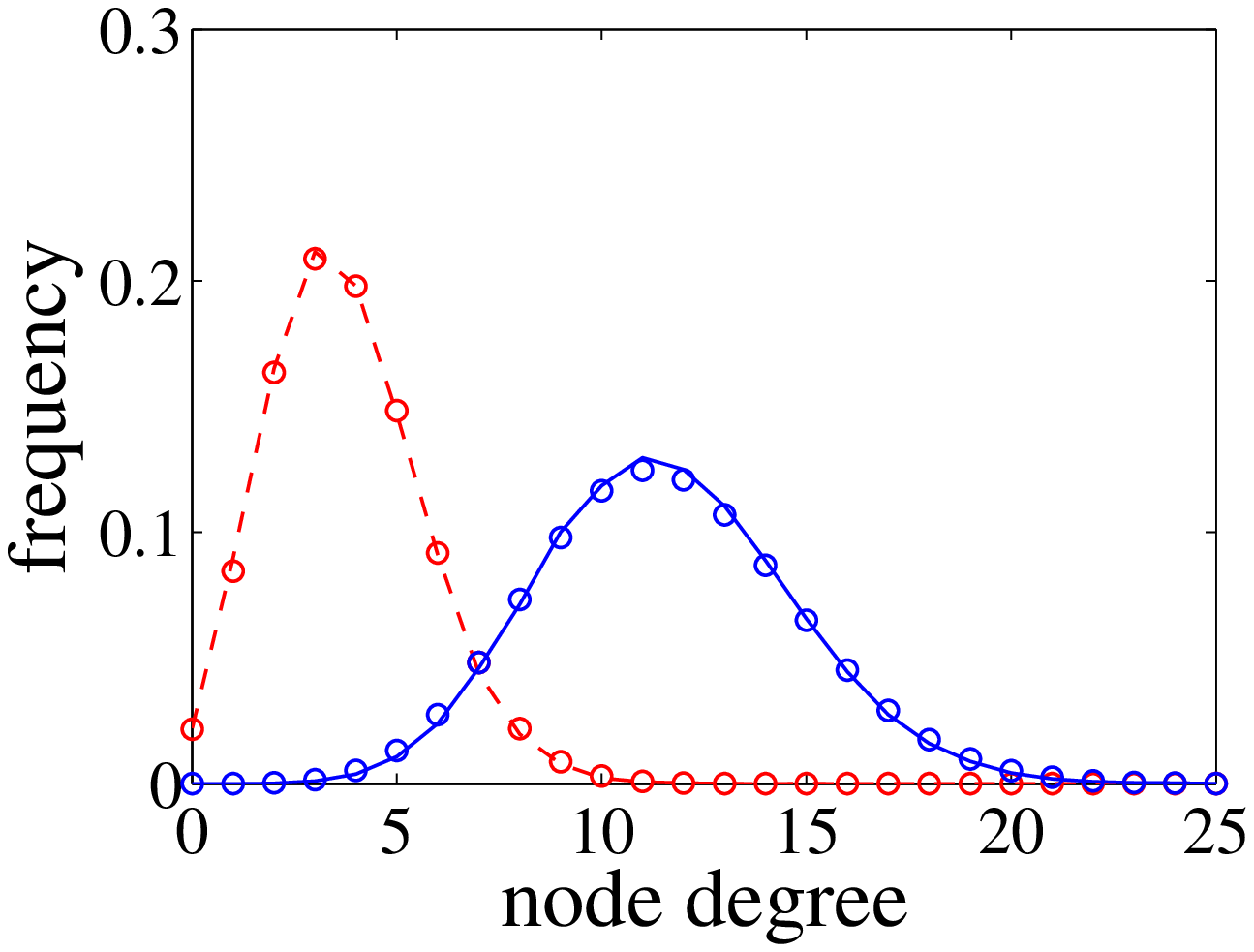}
    \caption{\label{fig-10k-distribution-label}Average degree distribution of all $S$ (blue line) and $I$ (red dashed line) nodes at the end of simulations, when starting from homogeneous (top) and heterogeneous (bottom) networks with node-based selection. The plots correspond to the average of 1000 simulations with $N$ = 100, $I_0$ = 20, $S_0 = N-I_0$, and $\langle k \rangle$ = 10. In the left panel, $R$ = $\sqrt{2N}/2$. In the right panel, $R$ = $\sqrt{6/\pi}$. The blue and red ($\star$) markers correspond to Eq.~\ref{homo-dis-S} and Eq.~\ref{homo-dis-I}, respectively. The blue and red ($\circ$) markers correspond to Eq.~\ref{hete-dis-S} and Eq.~\ref{hete-dis-I}, respectively. We note that our analytic derivation needs the number of $SI$ links that have been cut by the end of the rewiring process. This is taken from the simulation.}
\end{center}
\end{figure}

Starting with the no-constraint scenario, $R$ =  $\sqrt{2N} / 2$, Fig.~\ref{fig-10k-distribution-label} (left panel) confirms that the network has split into two disconnected networks, where the mean degrees of
susceptible and infected nodes at the end of the simulations are given by $\langle k_S \rangle$ = $(1+i_0)\langle k \rangle$ and $\langle k_I \rangle$ = $i_0\langle k \rangle$ where $s_0+i_0=1$. This is true when starting from either homogeneous or heterogeneous networks. As expected, the degree of $S$ nodes can only increase, while the degree of $I$ nodes strictly decreases. Starting with a homogeneous network, there is no $S$ node with a degree less than $\langle k \rangle$, and the maximum degree of $I$ nodes is at most $\langle k \rangle$ because all nodes have the same initial degree $k$.
For both homogeneous and heterogeneous networks, there are disconnected $I$ nodes at the end of the simulation, but, as discussed previously, this may result from the fact that $\langle k \rangle$ is not very
high.

\begin{figure}[h!]
\begin{center}
    \includegraphics[scale=0.28]{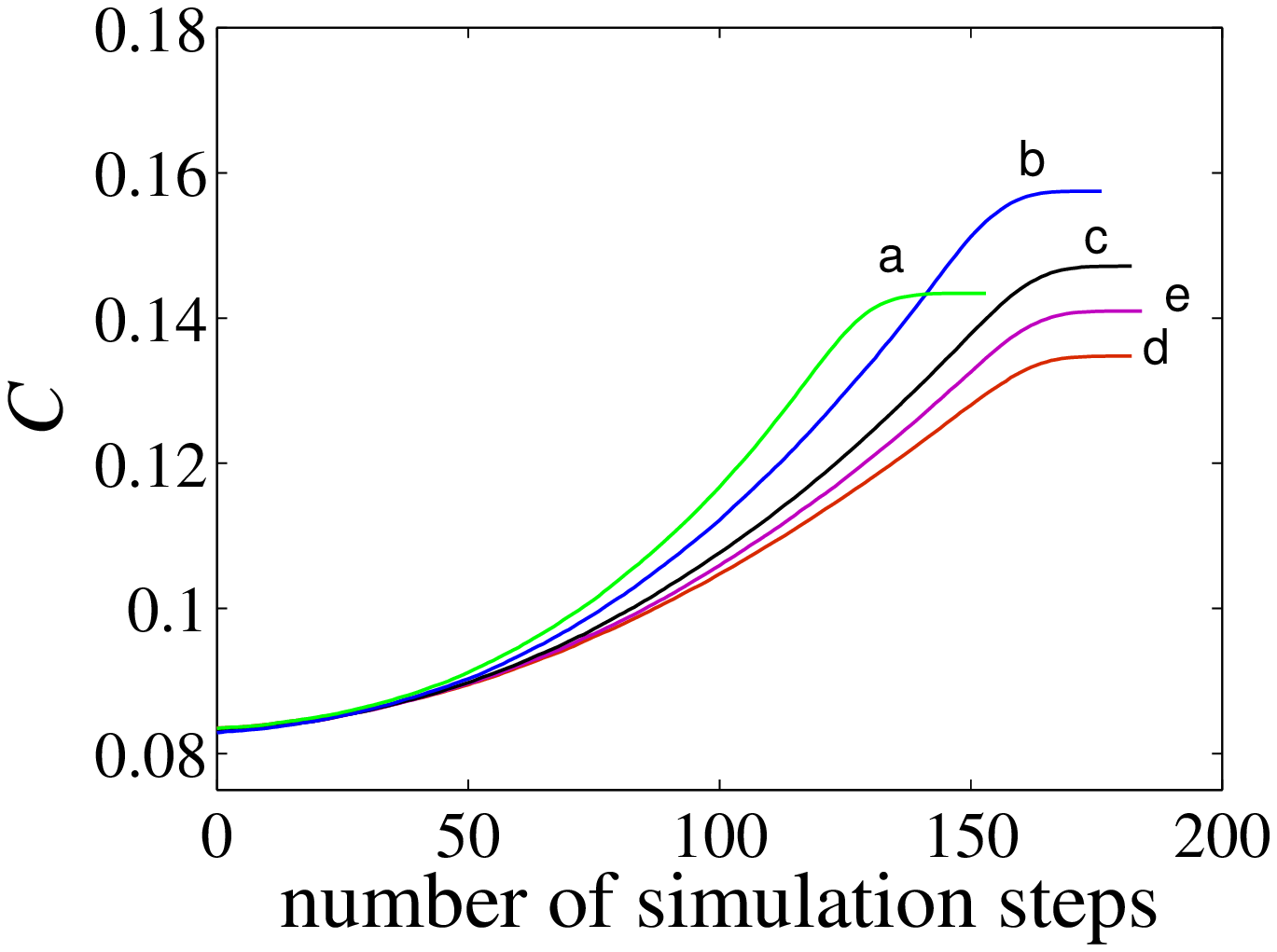}
    \includegraphics[scale=0.28]{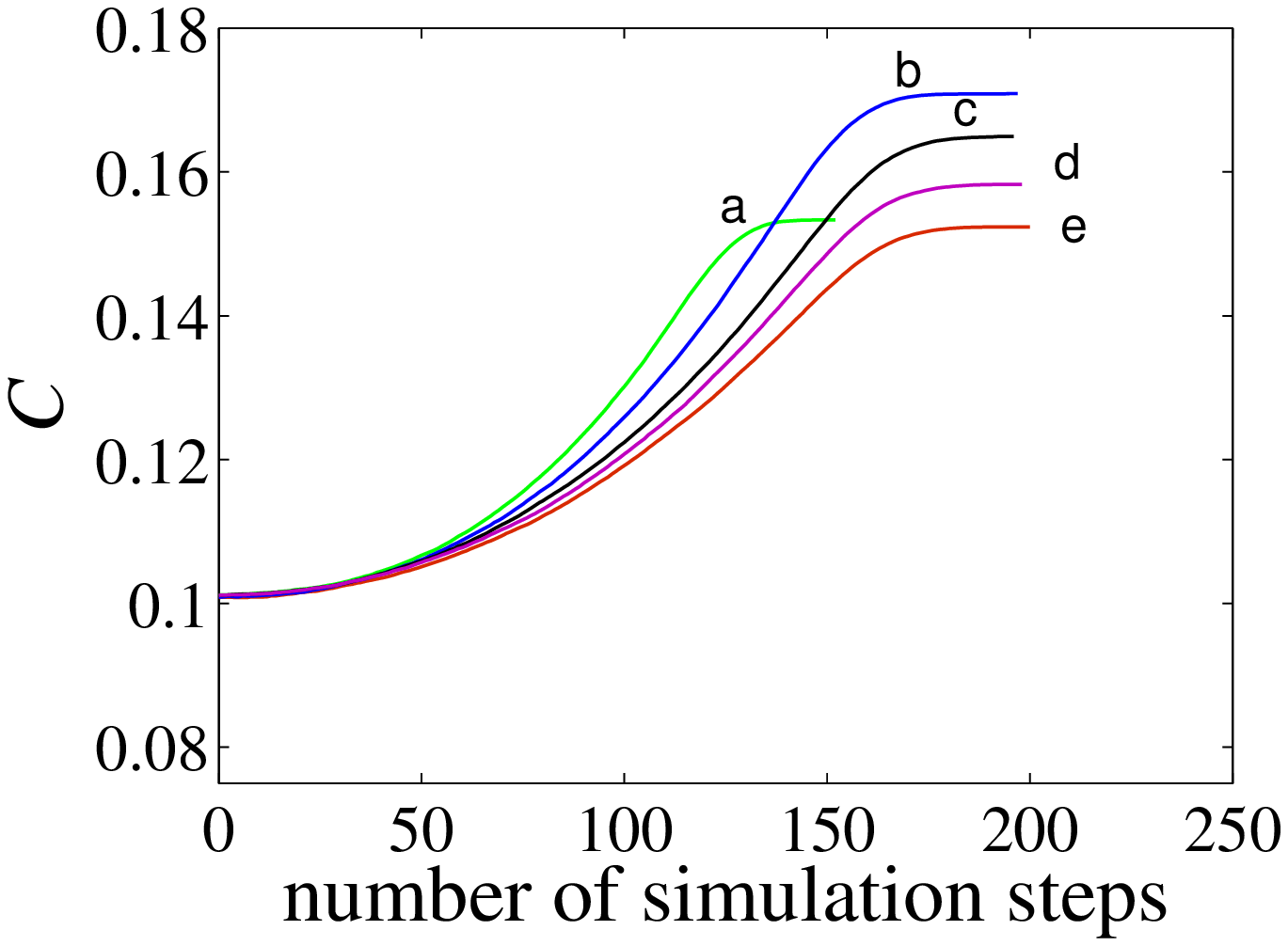}
    \caption{\label{fig-10k-clu-label}Evolution of clustering during rewiring, starting from homogeneous (left) and heterogeneous (right) networks. The plots correspond to the average of 1000 simulations with $N$ = 100, $I_0$ = 20, $S_0 = N-I_0$, and $\langle k \rangle$ = 10. Data for $R$ values of $\sqrt{6/\pi}$, $\sqrt{10/\pi}$, $\sqrt{20/\pi}$, $\sqrt{30/\pi}$ and $\sqrt{2N}/2$ are shown in green (a), blue (b), black (c), purple (d) and red (e), respectively.}
\end{center}
\end{figure}

For small local areas, e.g., $R$ =  $\sqrt{6/\pi}$ where the average number of nodes in a local area is smaller than the average degree, the rewiring is restricted by the limited number of available $S$ nodes. Therefore, the network evolves quickly to a stable equilibrium. This is clearly shown in Fig.~\ref{fig-10k-clu-label} in which the evolution of clustering for $R$ =  $\sqrt{6/\pi}$ stops (due to all rewiring being complete) before that of other (larger) radii $R$.

\begin{figure}[h!]
\begin{center}
    \includegraphics[scale=0.28]{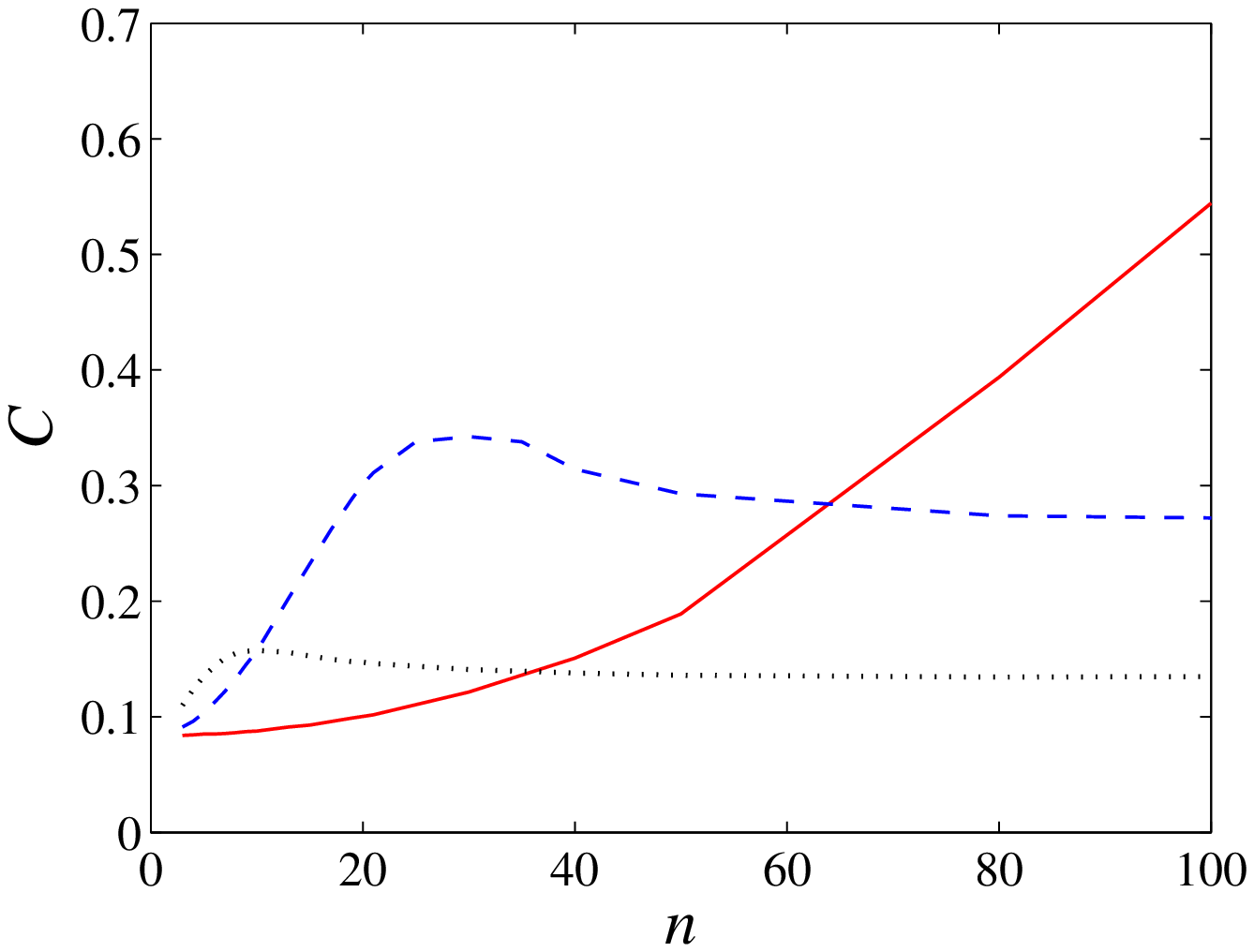}
    \includegraphics[scale=0.28]{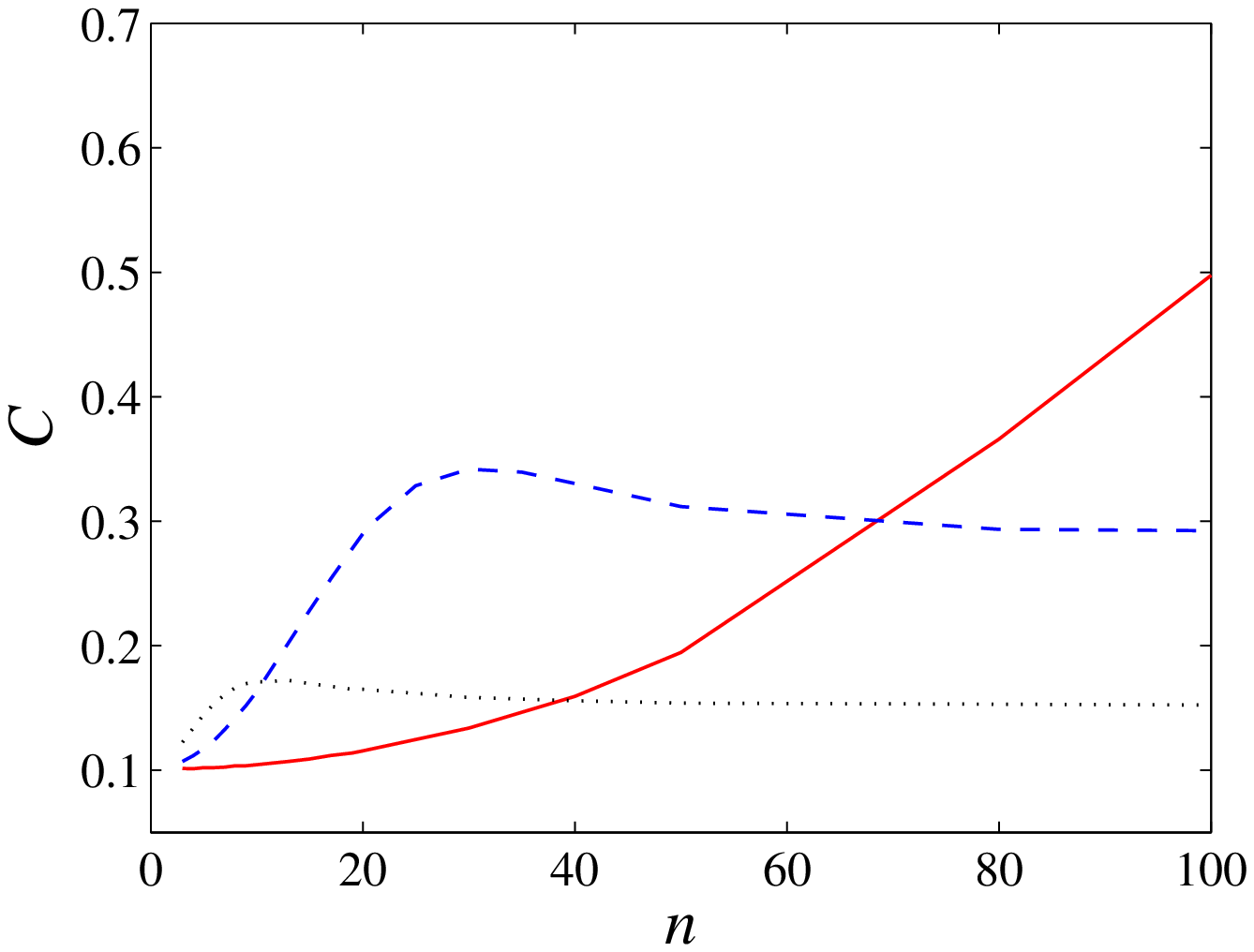}
    \caption{\label{fig-10k-clu}Final value of clustering when starting from homogeneous (left) and heterogeneous (right) networks. The plots correspond to the average of 1000 simulations with $N$ = 100, and $\langle k \rangle$ = 10. Data are shown for $I_0 = 20$ (black dotted line), $I_0 = 50$ (blue dashed line), $I_0 = 80$ (red solid line) with $S_0 = N-I_0$.}
\end{center}
\end{figure}

These results are not solely dependent on the spatial constraint, but also on the number of initial $SI$ links. Fig.~\ref{fig-10k-clu} shows the clustering at the end of the simulations for a range of radii $R$ and $I_0$ values. Starting with either homogeneous or heterogeneous networks produces similar results in clustering for a variety of $R$ and $I_0$ values. As expected, the maximum clustering values for all sets of parameters, $n$ and $I_0$, are not higher than the maximum clustering value for networks with no node labelling, obtained previously (see Fig~\ref{fig-clust-formulas}). A small number of initial $S$ nodes leads to a small number of successful rewiring events (see Fig.~\ref{fig-10k-clu} where $I_0=80$). This means that a larger value of $R$ is needed in order to find available $S$ nodes before cutting $SI$ links, and therefore, we find that clustering increases as the value of $R$ grows larger.

\section{\label{sec:full-model}$SIS$ models with constrained rewiring}
In the previous section, we showed that the spatially constrained rewiring plays an essential role in determining network structure in the absence of any node dynamics. In this section,
we extend this work by combining dynamics of the network with dynamics on the network in the form of the simple $SIS$ model. The simulations are carried out on both homogeneous and heterogeneous networks, with a fixed size of $N$ nodes and average degree of $\langle k \rangle$ links per node. The epidemic dynamics is specified in terms of infection and recovery events. The rate of transmission across an $SI$ link is denoted by $\tau$. Infected individuals recover independently of each other at rate $\gamma$. The network dynamics is specified in terms of rewiring events which affect $SI$ links. Here, we make the assumption that the rewiring of a $SI$ link depends on the number of susceptible nodes available for rewiring in the local neighbourhood of the $S$ node that wishes to break its link to an $I$ node and rewire to a susceptible one. It is natural to assume that the rewiring rate is proportional to the number of available $S$ nodes that can accept new connections. For all $SI$ links, this is achieved by using a rewiring rate equal to $h w$, where $h$ is the number of available susceptible nodes within $S$'s local area. We also assume that all processes are independent Poisson processes.

Simulations rely on synchronous updating with a small time step, $\Delta t$, which guarantees that at most one event happens per iteration. Only three different types of event are possible during one time step $\Delta t$: (a) infection of a susceptible $S$ node can occur with probability $1-exp(-k \tau \Delta t)$, where $k$ is the number of $I$ neighbours, (b) an infectious $I$ node
recovers with probability $1-exp(-\gamma \Delta t)$, and (c) a $SI$ link is rewired with probability $1-exp(-h w \Delta t)$, as long as $h>0$. This guarantees that rewiring only happens if viable candidates for rewiring exist and that the number of links in the network is constant throughout the simulation.

Given that the main focus of our study is the role of the spatially constrained rewiring, we will investigate the impact of the
$R$ (or $n$) values on whether epidemics die out and/or the endemic state becomes established.
Specifically, we use the following definition to characterise the impact of the expected number of nodes in a local area or size of local area:

\begin{defn}
$n^*$ is the critical value of the expected number of nodes in a circle-like local area such that any greater value of $n$ leads to disease extinction before a time $T$, or the end of the simulation, whichever comes first.
\end{defn}
%
%

\begin{figure}[h!]
\begin{center}
    \includegraphics[scale=0.28]{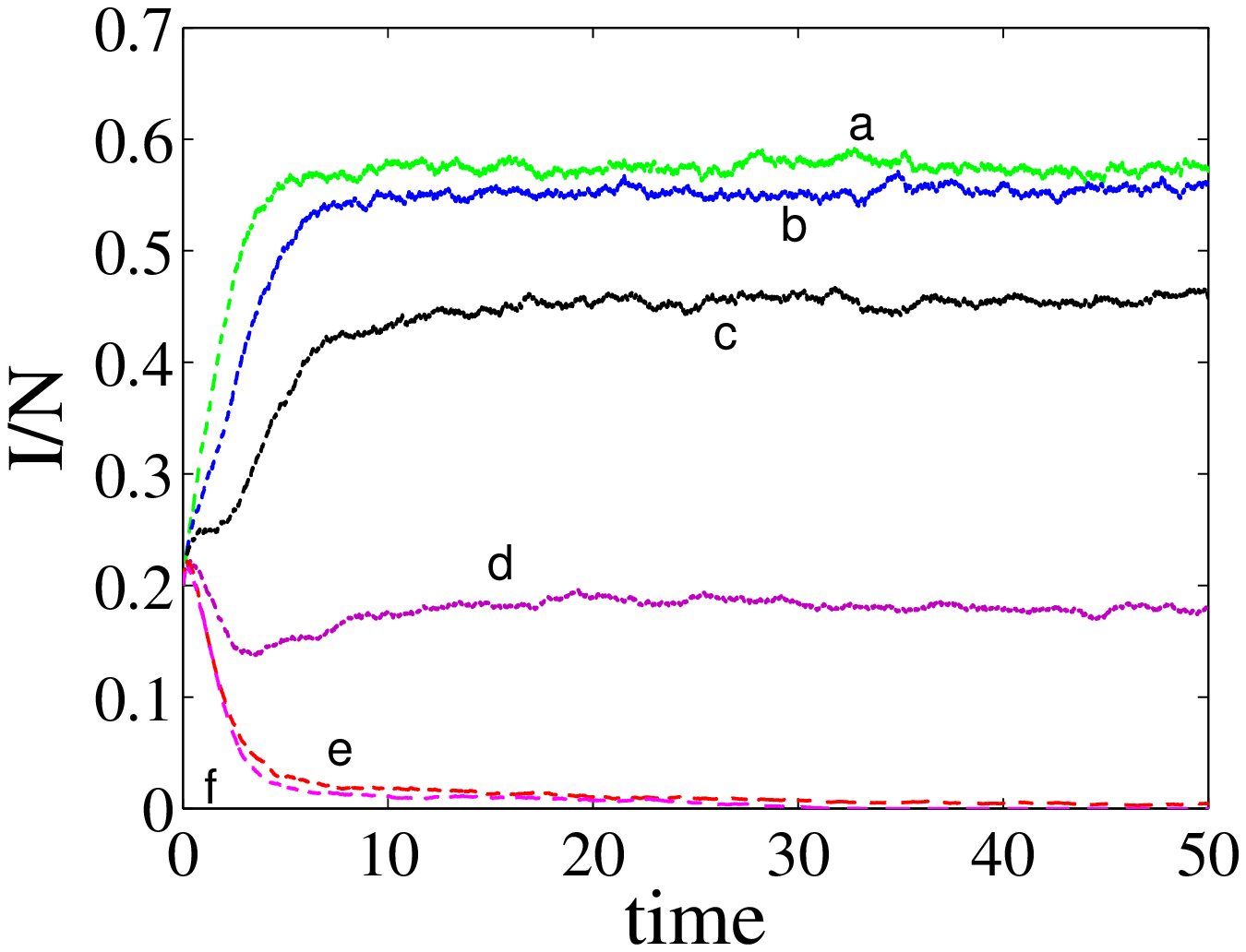}
    \includegraphics[scale=0.28]{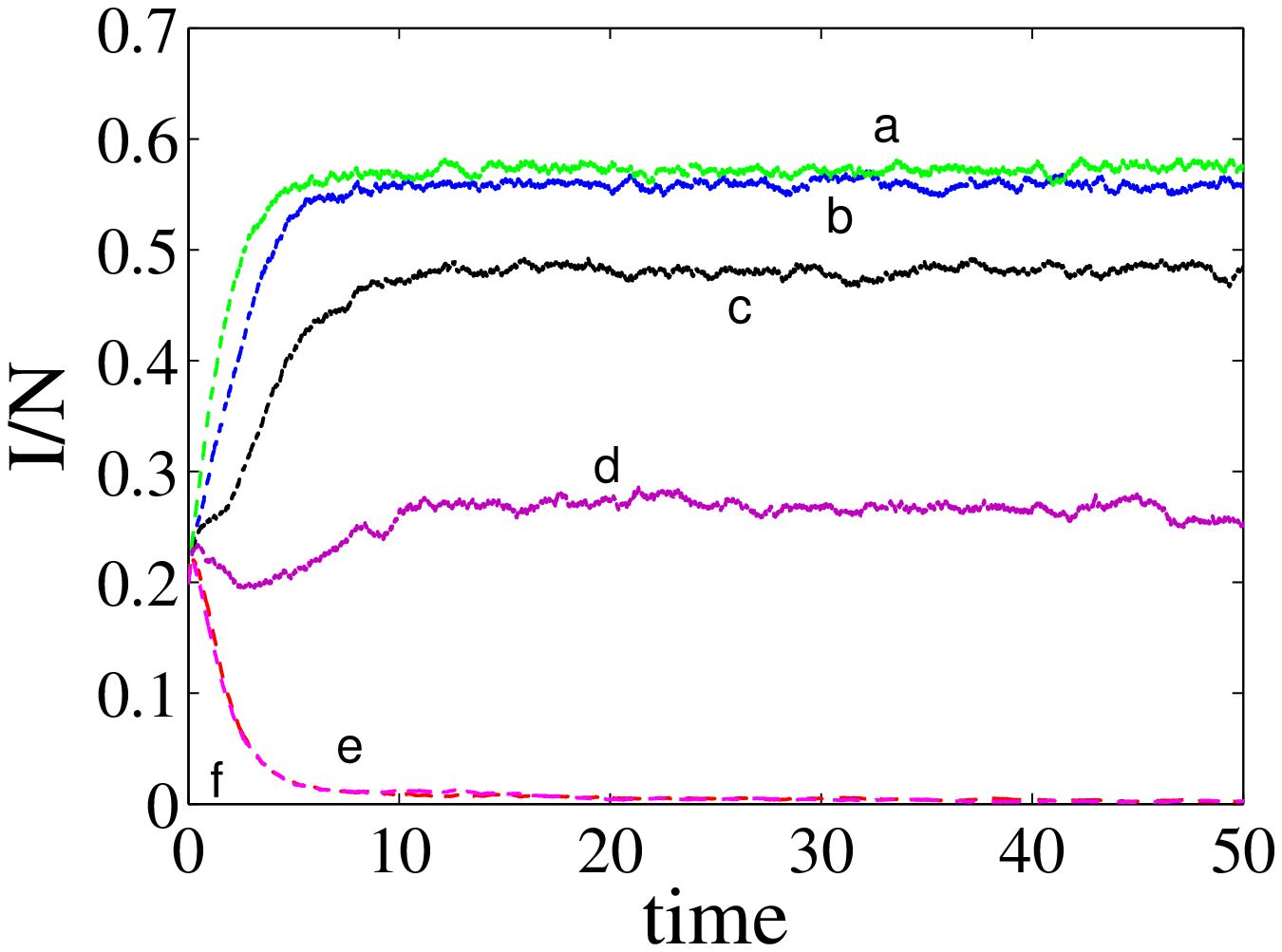}
    \caption{\label{fig-I}Infection prevalence ($I/N$) starting from homogeneous (left) and heterogeneous (right) networks. The plots correspond the average of 200 simulations with $N$ = 100, $I_0 = 20$, $S_0 = N-I_0$, $\langle k \rangle$ = 10, $\gamma=1$, $\tau=0.25$, $w=0.2$. Data are shown for $n$ values of 5 (green - a), 10 (blue - b), 15 (black - c), 20 (purple - d), critical value $n^{*}$ = 26 for homogeneous network and $n$ = 27 (red - e and pink - f, left panel), and, critical value $n^{*}$ = 29 and $n$ = 30 (red - e and pink - f, right panel)}.
\end{center}
\end{figure}

The time evolution of infection on adaptive networks with constrained rewiring is shown in Fig.~\ref{fig-I}. Here, all simulations use the following parameter values: $N = 100$, $\langle k \rangle = 10$, $\gamma = 1$ and final simulation time $T=100$.  Simulations are started with infectious nodes chosen at random. The controlling effect of the local area radius $R$ or expected number of nodes in a local area $n$ is clear to see. As expected, with a small value of $n$, the network dynamics does not play a significant role in the control of epidemic spread for either homogeneous or heterogeneous networks. The small value of $n$ affects the network dynamics in that the rewiring process can only happen briefly at the outset of the simulation and then stops while the epidemic dynamics continues throughout the simulation.

Larger values of $n$, however, creates ideal conditions for rewiring and this can continue throughout the simulation, resulting in breaking many $SI$ links. This scenario leads to a slowing down of the spread of the epidemic and a reduced infection prevalence. This is confirmed by Fig.~\ref{fig-I}, which shows small levels of infection prevalence for $n=15$ and $n=20$. The same figure also shows smaller and smaller endemic levels when the rewiring radius passes through the critical expected number $n^{*}$, namely, $n^{*} = 26$ for homogeneous networks and $n^{*} = 29$ for heterogeneous networks.

\begin{figure}[h!]
\begin{center}
    \includegraphics[scale=0.28]{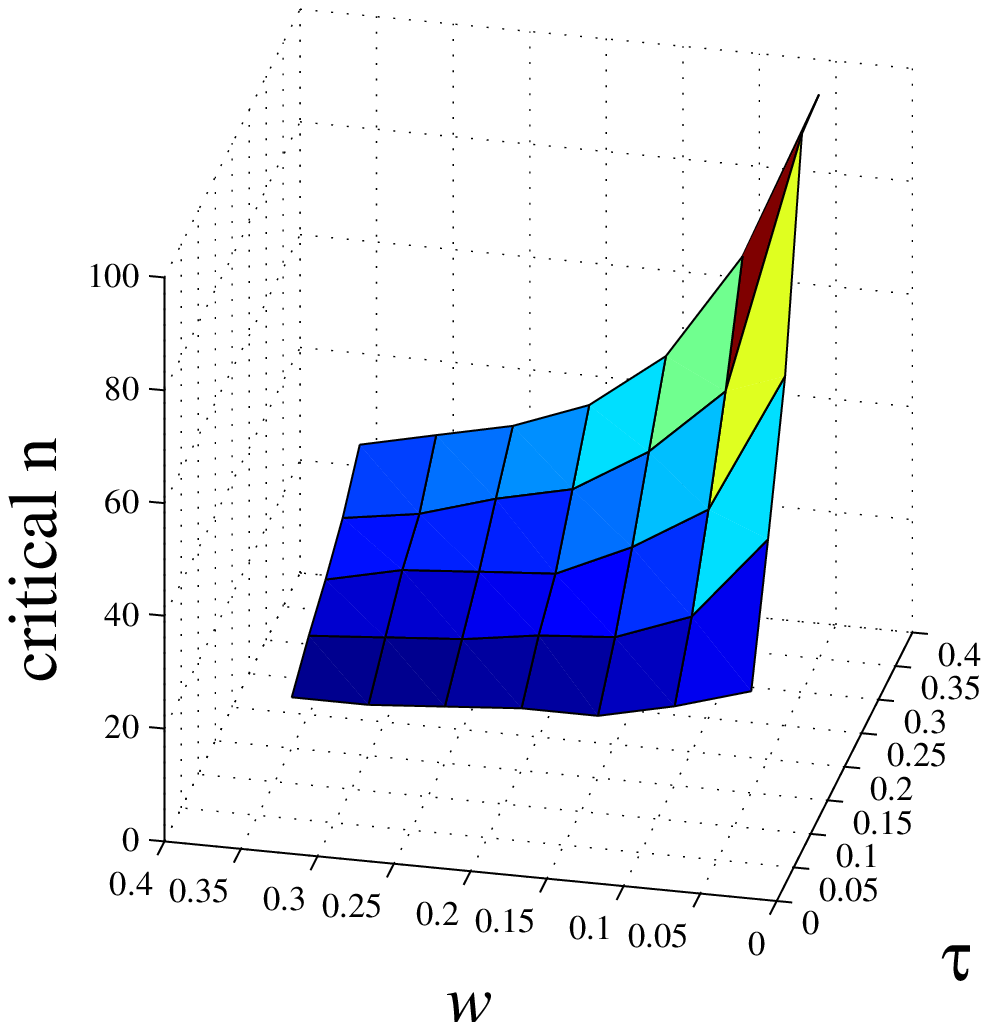}
    \includegraphics[scale=0.28]{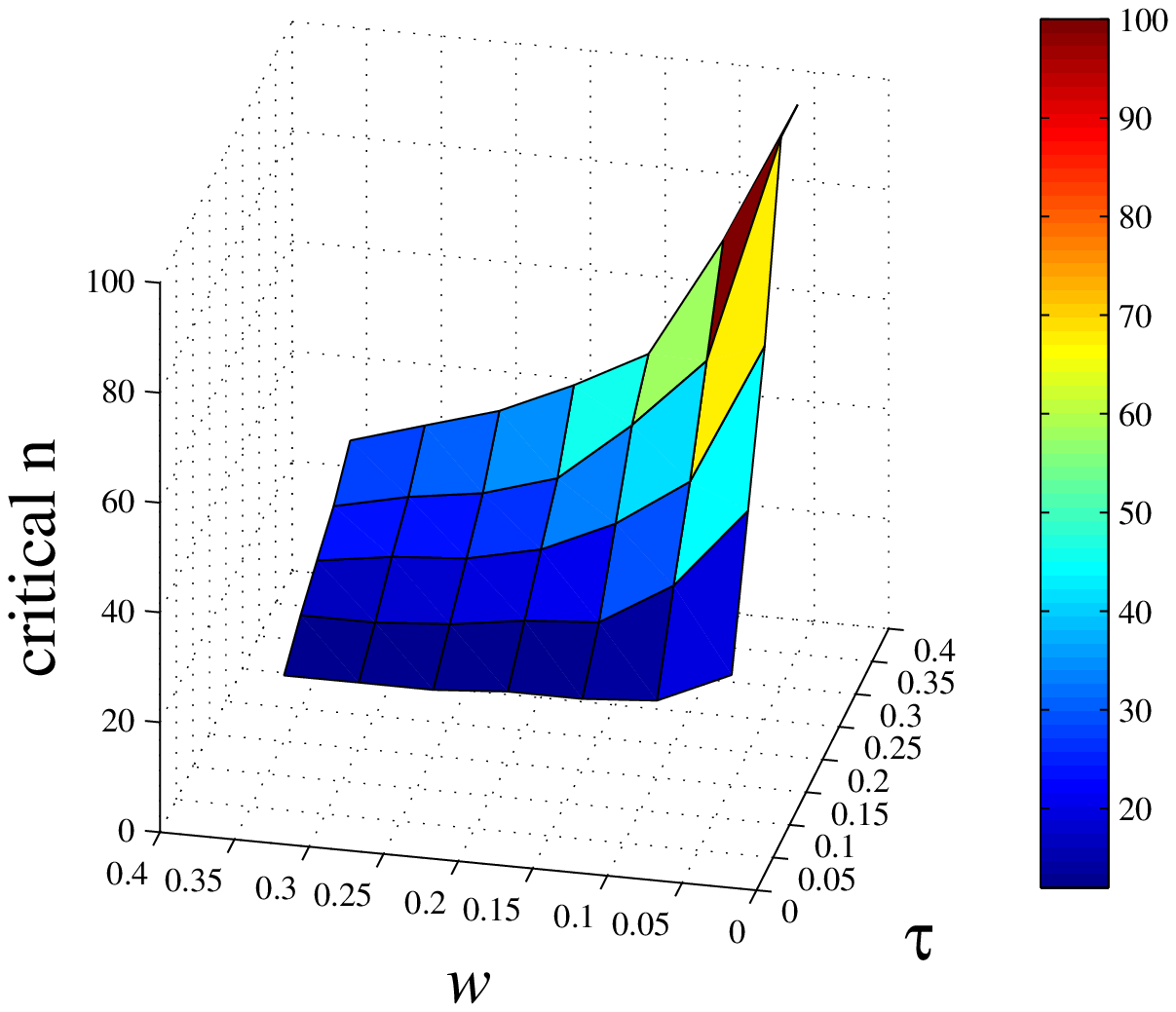}
    \caption{\label{fig-nR-critical}Critical $n$ as a function of $\tau$ and $w$, starting from homogeneous (left) and heterogeneous (right) networks, and with $N$ = 100, $I_0 = 20$, $S_0 = N-I_0$, $\langle k \rangle$ = 10 and $\gamma=1$.}
\end{center}
\end{figure}

To further understand the relationship between the critical value, $n^{*}$, and the disease parameters, we systematically varied the infectious and rewiring rates (with fixed recovery rate) and recorded the corresponding critical $n^{*}$ value. Fig.~\ref{fig-nR-critical} shows the resulting surface for both homogeneous and heterogeneous networks, where $\tau$ varies
from 0.15 to 3.5 in steps of 0.05 and $w$ varies from 0.05 to 0.35 in steps of 0.05.

\begin{figure}[h!]
\begin{center}
    \includegraphics[scale=0.28]{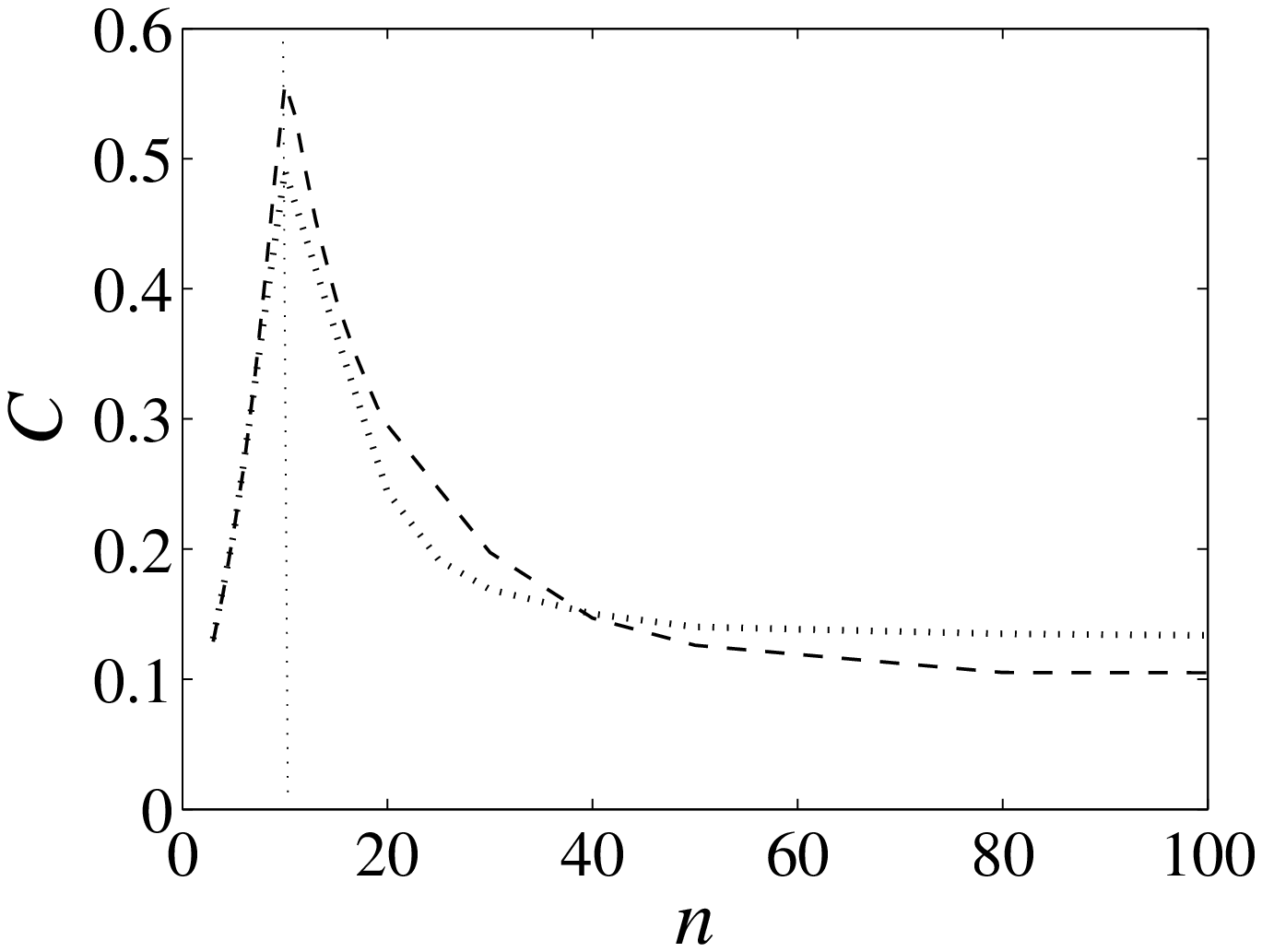}
    \includegraphics[scale=0.28]{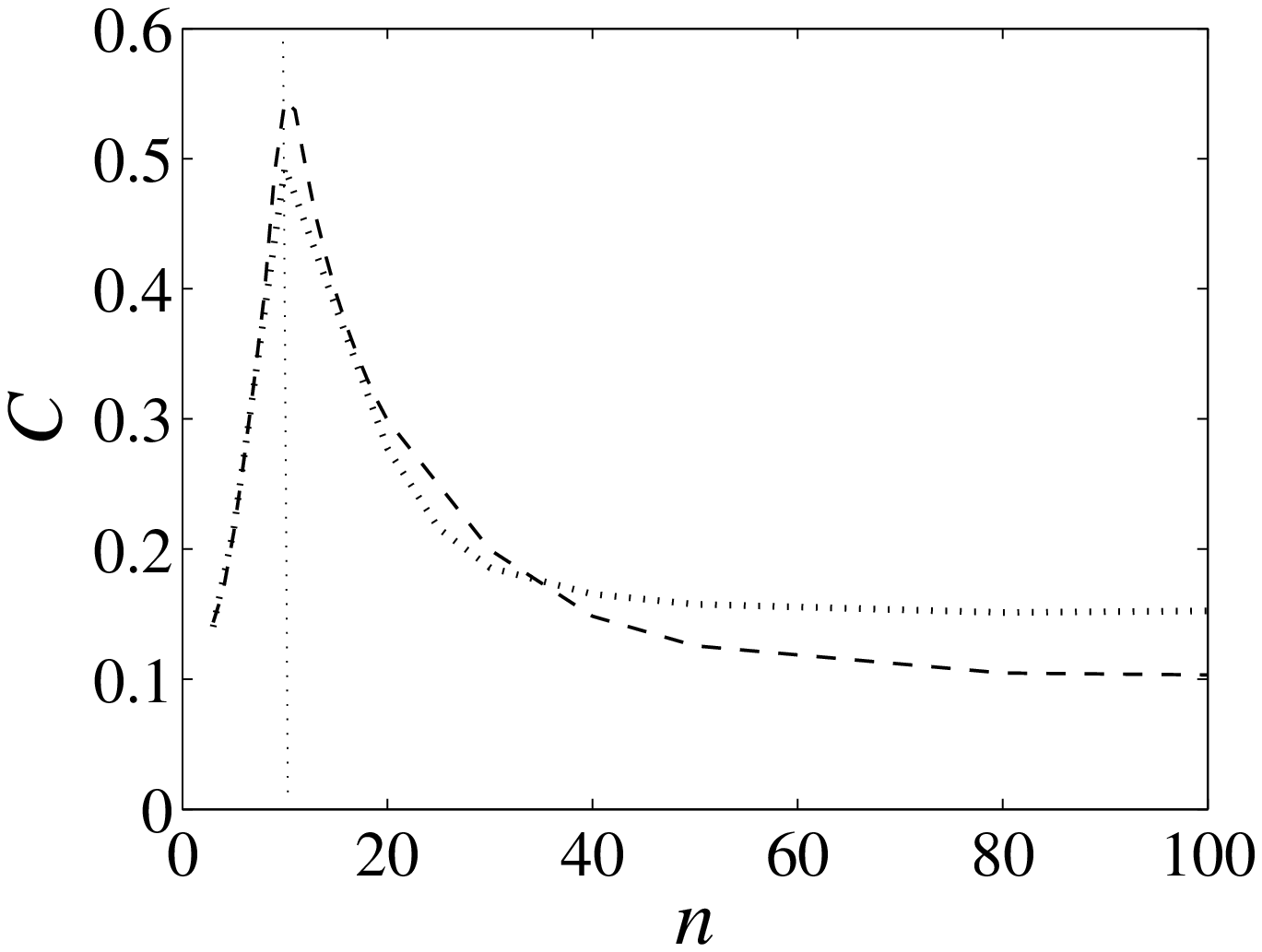}
    \caption{\label{fig-clustering-without-full}Final value of clustering starting from homogeneous (left) and heterogeneous (right) networks with $N$ = 100 and $\langle k \rangle$ = 10. The dashed line shows the clustering in a network without any dynamics of the nodes and without node labelling. The dotted line denotes the clustering when the full model, couple epidemic dynamics and rewiring, is considered, with $I_0 = 20$, $S_0 = N-I_0$, $\gamma=1$, $\tau=0.25$ and $w=0.2$.}
\end{center}
\end{figure}

Increasing values of $n$ increase the rewiring rate, $hw$, since $h$ will be higher due to more targets for the rewiring being available. This in turn leads to an active rewiring process which results in an overall decrease in the endemic equilibrium or in the extinction of the epidemic.

It is found that when $n$ is large, the starting configuration of the network affects the endemic equilibrium in so far as starting with a homogeneous network leads to a smaller epidemic, at a given $n$, than when starting with a heterogeneous network. Typically, the critical $n^{*}$ is higher for heterogeneous networks, meaning that rewiring needs to be less constrained in order to curtail the epidemic. In general, for all $n$ values, the epidemic will spread faster on heterogenous networks early on in the epidemic, when the link rewiring is still limited. However, as the networks are rewired, this effect is weakened as the homogenous network will become more heterogenous and will become more similar to the networks started with heterogenous degree distributions. Nevertheless, the critical threshold differs between homogenous and heterogenous networks, which may reflect  a build up of structural correlations or differences which may differentially affect the endemic prevalence.

In Fig.~\ref{fig-clustering-without-full}, we present the final clustering value for a range of radii $R$ for both the full model and the model with no epidemic or labelling. The simulation results show that the impact of changing the radius on network structure is similar in both cases. Specifically, high values of $n$, but with $\langle k \rangle \gg n$ (the region to the left of the vertical line), result in higher levels of clustering, whereas when $n$ is such that $\langle k \rangle \ll n$ (see the region to the right of the vertical line), clustering decreases, irrespective of which network is used.
It is worth noting that the analysis of the dynamic network model alone, without labelling or epidemic, gives a clear insight into how the structure of the network changes.
Observations from this analysis still hold in the full model, but as expected, the clustering of networks in the full model is less than in the network-only model since labelling reduces the
number of nodes that can be used when rewiring. Higher clustering values in the full model are due to the epidemics dying out quickly with no further rewiring, and thus with the network
displaying a clustering value that is close to the values observed in the starting network. For the network only model or for full blown epidemics, however, the network will be fully randomised.

\section{\label{sec:discussion}Discussion}
The present study explored the effect of spatially constrained rewiring on an $SIS$ epidemic unfolding on an adaptive network. Specifically, the dynamics of the network was achieved by breaking links and reconnecting to nodes within a local area. A step-by-step approach was taken in which the network dynamics was studied first in the absence of disease dynamics, then with node labelling but no dynamics, and finally with both network and node dynamics. Two different starting networks were used and analysed. In all models, a range of radii $R$, giving circular neighbourhoods, within which to rewire, was considered and shown to provide the means to control epidemic outbreaks. Spatially limited rewiring provides a more realistic mechanism than choosing partners to rewire to from the entire population. It is highly likely that in most situations, rewiring will be limited to a limited sub-populations or set of individuals.

Our study provided a detailed analysis of the impact of constrained rewiring on the structure of the network. In particular, we were able to give analytic and semi-analytic results for degree-distribution and clustering. These showed excellent agreement with simulations and we have revealed that it is possible to generate networks with the same mean path length, same clustering but significantly different distribution of real link lengths. This comes in support to the findings of~\cite{ritchieJTB2014} that networks with same clustering can have substantially different higher-order network structure. This needs further investigation, possibly using more complex node dynamics to reveal how subtle differences in the network structure may impact on the outcome of dynamical processes supported by the network.

Further results provided analytical formulas for the degree distributions of susceptible and infected nodes which again showed good agreement with simulation results. These also confirmed that starting from a heterogeneous network, and when $R$ is equal to $\sqrt{2N}/2$ or in the absence of spatial constraints for rewiring, the average degree of $S$ and $I$ nodes are $(1+i_0) \langle k \rangle$ and $i_0 \langle k \rangle $, respectively, which is in line with~\cite{Gross2006}.

Finally, we have shown that even constrained rewiring can serve as a potent control measure.
We highlighted that the expected number $n$ in a typical local area is a key parameter which influences
the network dynamics and can determine whether disease dies out or becomes endemic.
Extensions to the methodology presented in this paper include considering other forms of constrained rewiring, e.g., network models where locality is not just defined in terms of spatial distance but possibly some more abstract or general metric, and understanding how this impacts network structure and on processes, other than epidemics, taking place on the network.

\begin{acknowledgments}
P. Rattana acknowledges funding for her Ph.D. studies from the
Ministry of Science and Technology, Thailand.
\end{acknowledgments}

\bibliography{apstemplate_paper_reprintformat}

\end{document}